\documentclass[twocolumn]{aastex62}

\graphicspath{{./}{figures/}}

\received{July 28, 2021}
\revised{September 20, 2021}
\accepted{October 5, 2021}

\submitjournal{ApJ}
\shorttitle{NTF Magnetism}
\shortauthors{Par\'e et al.}

\begin{document}

\title{Analyzing the Intrinsic Magnetic Field in the Galactic Center Radio Arc}

\correspondingauthor{Dylan Par\'e}
\email{dylan-pare@uiowa.edu}

\author[0000-0002-5811-0136]{Dylan M. Par\'e}\footnote{Also a CSIRO Space and Astronomy CASS student}
\affil{University of Iowa \\
30 North Dubuque Street, Room 203 \\
Iowa City, IA 52242}

\author[0000-0002-7491-7386]{Cormac R. Purcell}
\affil{Department of Physics and Astronomy \\
Macquarie University \\
NSW 2109 Sydney, Australia}

\author{Cornelia C. Lang}
\affil{University of Iowa \\
30 North Dubuque Street, Room 203 \\
Iowa City, IA 52242}

\author[0000-0002-6753-2066]{Mark R. Morris}
\affil{University of California, Los Angeles \\
430 Portola Plaza, Box 951547 \\
Los Angeles, CA 90095-1547}

\author[0000-0002-2670-188X]{James A. Green}
\affil{Commonwealth Scientific and Industrial Research Organization \\
Space and Astronomy\\
26 Dick Perry Avenue \\ 
Kensington, WA 6151}

\begin{abstract}

The Radio Arc is a system of organized non-thermal filaments (NTFs) located within the Galactic Center (GC) region of the Milky Way. Recent observations of the Radio Arc NTFs revealed a magnetic field which alternates between being parallel and rotated with respect to the orientation of the filaments. This pattern is in stark contrast to the predominantly parallel magnetic field orientations observed in other GC NTFs. To help elucidate the origin of this pattern, we analyze spectro-polarimetric data of the Radio Arc NTFs using an Australian Telescope Compact Array data set covering the continuous frequency range from $\sim$4 to 11 GHz at a spectral resolution of 2 MHz. We fit depolarization models to the spectral polarization data to characterize Faraday effects along the line-of-sight. We assess whether structures local to the Radio Arc NTFs may contribute to the unusual magnetic field orientation. External Faraday effects are identified as the most likely origin of the rotation observed for the Radio Arc NTFs; however, internal Faraday effects are also found to be likely in regions of parallel magnetic field. The increased likelihood of internal Faraday effects in parallel magnetic field regions may be attributed to the effects of structures local to the GC. One such structure could be the Radio Shell local to the Radio Arc NTFs. Future studies are needed to determine whether this alternating magnetic field pattern is present in other multi-stranded NTFs, or is a unique property resulting from the complex interstellar region local to the Radio Arc NTFs.
\end{abstract}


\section{INTRODUCTION} \label{sec:intro}

The Galactic Center (GC) is the nearest galactic nuclear region to Earth, being only 8.0 kpc away \citep{Abuter2019,Do2019}. Its proximity allows us to infer properties of more distant galactic nuclear regions if we assume that the Milky Way is a representative spiral galaxy. The GC contains a population of unique structures that appear as long, illuminated threads at radio frequencies \citep{Gray1995,Morris1996,Yusef-Zadeh2004}.

These structures are known as the non-thermal filaments (NTFs) and are highly polarized, synchrotron sources \citep{YMC1984,YM1987}. The NTFs are generally an order of magnitude or more longer than they are wide \citep{Morris2007}. Their synchrotron nature indicates the presence of relativistic, free electrons \citep{YMC1984}. The source and mechanism causing the electrons to accelerate to relativistic velocities, however, remain a topic of active debate with many hypotheses (see references in \citealt{Ponti2021}).

The first observed NTF system is also the most prominent, and is known as the GC Radio Arc (hereafter referred to as the ``Radio Arc NTFs''), consisting of $\rm\geq$10 individual filaments \citep{Yusef-Zadeh1986a,Yusef-Zadeh1987}. These filaments are generally parallel to one another and are oriented roughly perpendicular to the Galactic plane. The individual filaments are quite narrow, with narrowest widths of $\rm\sim$0.5'' (or 0.02 pc) \citep{Pare2019}. The strength and striking morphology of the total intensity emission of the Radio Arc NTFs make it a compelling target for probing the nature of the larger NTF population.

In general, the NTFs are found to have discontinuous polarized intensity distributions in comparison to their continuous total intensity distributions \citep{Yusef-Zadeh1989,Gray1995,YWP1997,Lang1999a,Lang1999b}. Though the polarized intensity is discontinuous, it is seen to closely trace the total intensity distributions of the NTFs. While the polarized intensity does trace the total intensity of the Radio Arc NTFs \citep{Yusef-Zadeh1987,Inoue1989}, structures having a high fractional polarization are observed which extend into regions of low total intensity (e.g. \citealt{Pare2019}). 

\citet{Pare2019} also determined the magnetic field for the Radio Arc NTFs for the first time by attempting to correct for Faraday rotation along the line-of-sight by linear fitting the polarization angle as a function of $\rm\lambda^2$ to determine the amount of Faraday rotation along each line of sight. They found that the magnetic field of the Radio Arc NTFs alternated from being parallel to rotated with respect to the orientation of the NTFs. In contrast, the magnetic fields seen for other NTFs have been predominantly parallel to the NTFs \citep{YM1987,Gray1995,YWP1997,Lang1999a,Lang1999b}. In this paper, we use the term ``parallel magnetic field'' to indicate a magnetic field that is oriented along the NTFs, whereas ``rotated magnetic field'' refers to a magnetic field oriented at an angle rotated from the NTFs.

\citet{Pare2019} explained the unusual polarized intensity features and the rotated magnetic field regions they observe as being components of an intervening structure local to the GC which was only partially resolved in their Very Large Array (VLA) \footnote{The National Radio Astronomy Observatory is a facility of the National Science Foundation operated under cooperative agreement by Associated Universities, Inc.} observations. They proposed that this structure could be the Radio Shell encompassing the observed portion of the Radio Arc NTFs \citep{Simpson2007,Pare2019}.

To build on the findings of \citet{Pare2019}, we have observed the Radio Arc NTFs using the Australia Telescope Compact Array (ATCA)\footnote{The Australia Telescope Compact Array is part of the Australia Telescope National Facility which is funded by the Australian Government for operation as a National Facility managed by CSIRO.} from 4-11 GHz using multiple configurations. These observations provide contiguous frequency coverage over a large frequency bandwidth and the various configurations used provide sensitivity to sources with a range of angular sizes. The contiguous frequency coverage provided by this data set allows us to employ sophisticated analysis techniques to more comprehensively study the arrangement of magnetized interstellar structures located along the line-of-sight toward the Radio Arc NTFs \citep{Brentjens2005,OSullivan2018}. In particular, we use our techniques to investigate the impact of the Radio Shell on the polarized emission of the Radio Arc NTFs.

In Section \ref{sec:data} we describe the observations and data reduction process. In Section \ref{sec:res} we describe the total intensity, polarized intensity, and RM distributions and outline the depolarization and rotation models considered for the fitting.  Section \ref{sec:model_res} details the results of our model fitting. Section \ref{sec:disc} discusses the results obtained from our model fitting and Section \ref{sec:conc} presents the conclusions of this work.

\section{OBSERVATIONS AND DATA REDUCTION} \label{sec:data}

\subsection{Observations} \label{sec:obs}
\begin{deluxetable*}{|c|c|c|c|c|}[ht!]
\tablecaption{Summary of ATCA Observations}
\tablecolumns{6}
\tablenum{1}
\tablewidth{0pt}
\tablehead{
\colhead{Freq. Range (GHz)} & \colhead{T$\rm_{int}$/field (min)} & \colhead{N$\rm_{fields}$} & \colhead{Array Configs.\tablenotemark{a}} & \colhead{Total time (hrs)}
}
\startdata
1.10 - 3.10 & 80 & 3 & 6a, 1.5d, 750c & 15.6 \\
4.00 - 6.00 & 40 & 7 & 6a, 1.5d, 750c, ew367 & 24.0 \\
5.75 - 7.74 & 40 & 7 & 6a, 1.5d, 750c, ew367 & 24.0 \\
7.50 - 9.50 & 40 & 7 & 6a, 1.5d, 750c, ew367 & 24.0 \\
8.80 - 10.80 & 40 & 7 & 6a, 1.5d, 750c, ew367 & 24.0 \\
\enddata
\tablecomments{Freq. Range shows the frequency range of each frequency band in GHz, T$\rm_{int}$/field shows the amount of time, in minutes observed for each pointing, N$\rm_{fields}$ shows the number of fields observed, Array Configs. details the ATCA arrays used for each frequency band, and Total time shows the total time observed for each frequency band in hours.}
\tablenotetext{a}{Configurations listed from largest (6\,km baselines) to smallest (367\,m baselines), the letter denotes a different configuration and `ew' is east-west (although all configurations used are purely in an east-west linear configuration).}
\end{deluxetable*}

The Radio Arc NTFs were observed with multiple ATCA array configurations using four different frequency bands to provide contiguous frequency coverage from 4.0$-$10.8 GHz, sampled every 2 MHz. Details of this data set are shown in Table 1. We also observed the Radio Arc NTFs at a lower frequency band from 1.1$-$3.1 GHz, which is also shown in Table 1; however, this frequency band suffered from severe Radio Frequency Interference (RFI) during our observations and more than 50\% of the channels were flagged due to this issue. Failure to account for RFI would impact the polarized intensity spectrum, by either producing noisy channels in the polarized intensity spectrum or forcing these channels to be removed from the spectrum. Noisy or missing channels in the polarized intensity spectra would decrease our sensitivity to Faraday media, making it more challenging to identify discrete depolarization mechanisms. 

\begin{figure}[hb!]
    \centering
    \includegraphics[width=8.5cm]{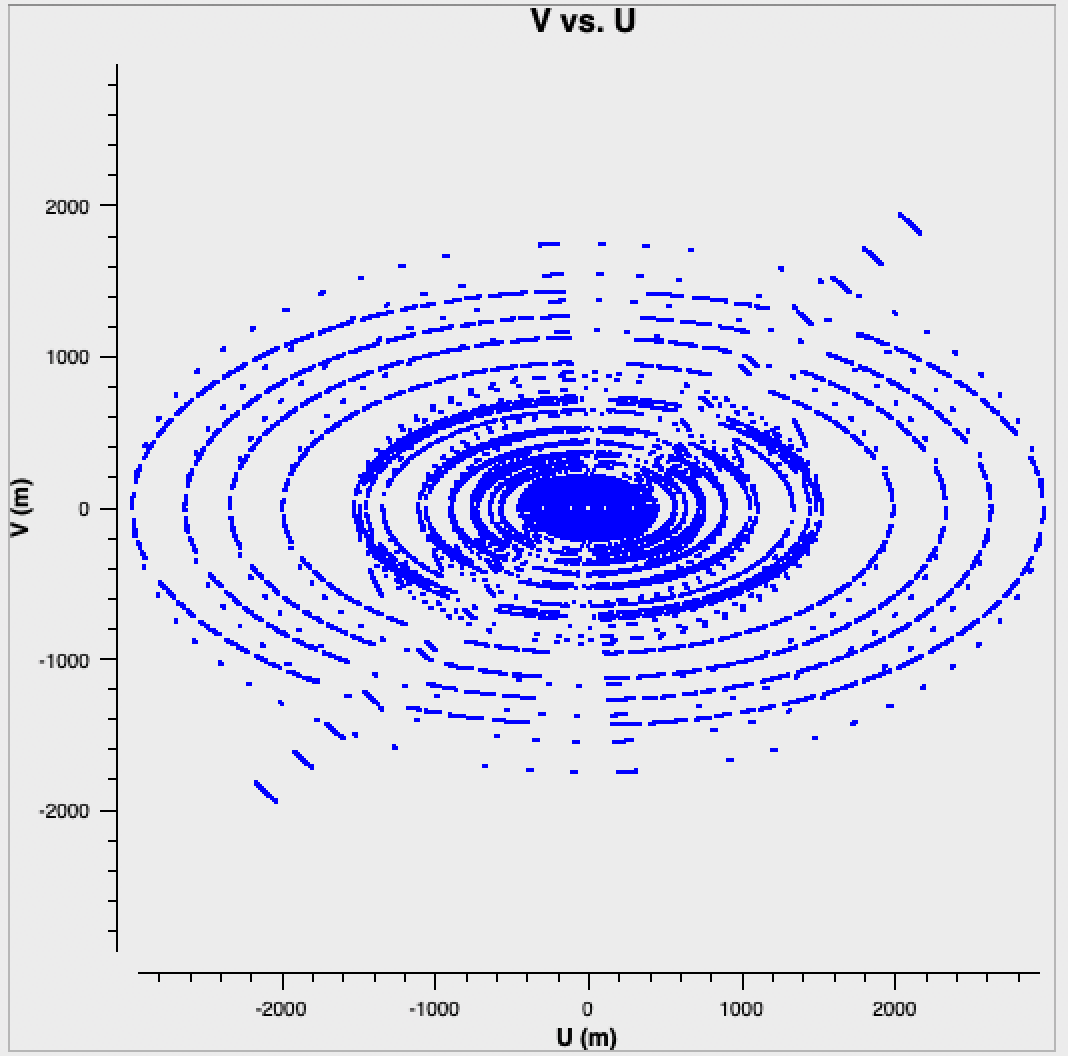}
    \includegraphics[width=8.5cm]{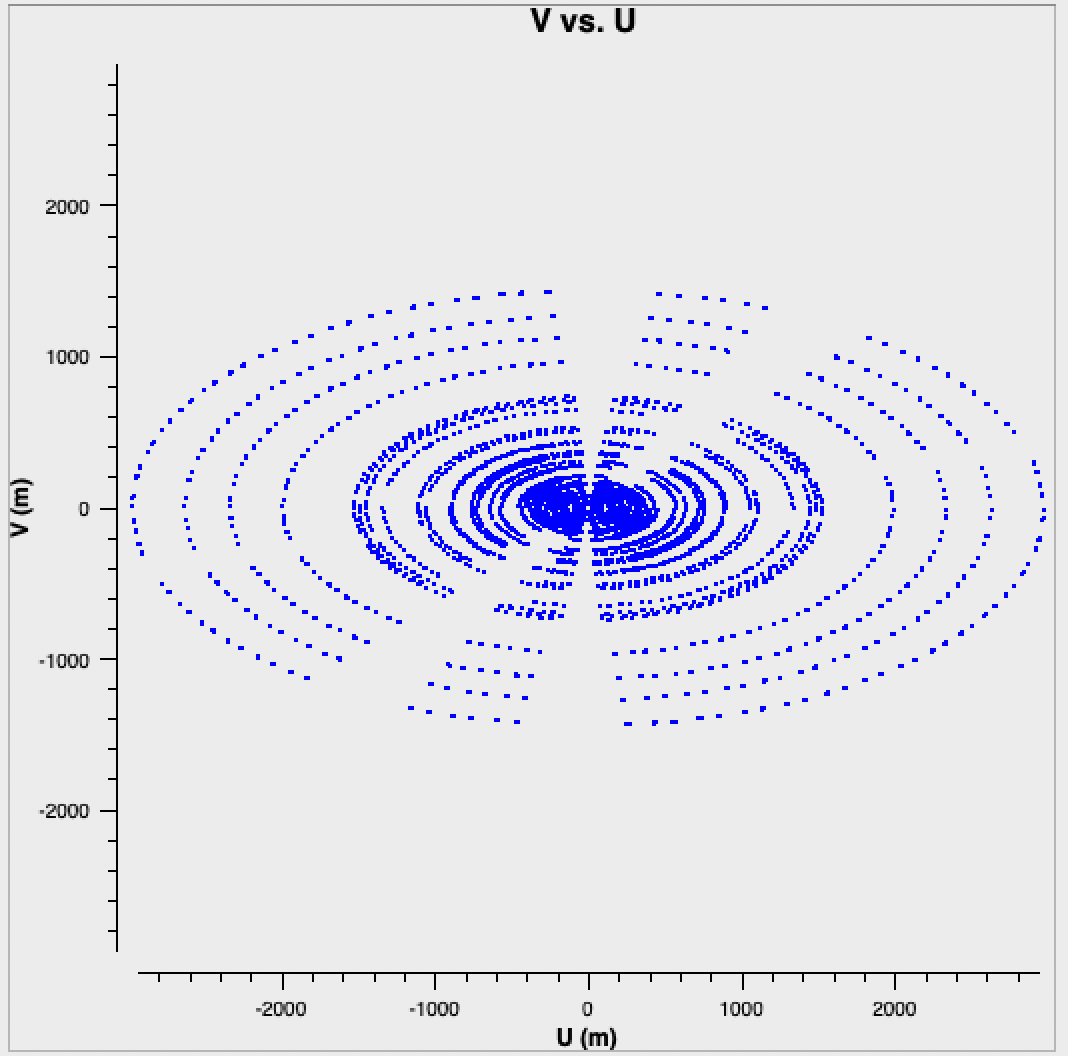}
    \caption{Representative uv plots at 5000 MHz for a single day of observation for all target and calibrator observations (upper) and for a single pointing of the Radio Arc NTFs (lower).}
    \label{fig:uv_samp}
\end{figure}

The comparatively small number of antennas of the ATCA telescope (six), coupled with the purely East-West linear configuration, provides limited instantaneous baseline coverage, but this is compensated by observing the Radio Arc NTFs for multi-hour observation blocks, allowing for more complete uv-coverage. Figure \ref{fig:uv_samp} shows a typical uv-coverage plot for a single frequency band as a result of these multi-hour observation blocks. The lower panel of Figure \ref{fig:uv_samp} shows the uv-coverage for a single pointing of the Radio Arc NTFs. In the upper panel of Figure \ref{fig:uv_samp} the line of uv-tracks extending at a 45 degree angle from the u-axis are the observations of calibrator PKS 1934-638. The isolated short-time uv-points spaced at regular intervals throughout the uv-plane in the upper panel of Figure \ref{fig:uv_samp} are the phase calibrator PKS 1827-360 observations. The remaining uv-tracks in this panel are for observations of the Radio Arc NTFs. 

Observations at all frequency bands except for the lowest frequency band consisted of seven target pointings with six observing different portions of the Radio Arc NTFs, and a seventh centered on Sgr A$\rm^*$. For the data from 1.1 - 3.1 GHz, only two pointings on the Radio Arc NTFs were necessary to image the full region, with a third centered on Sgr A$\rm^*$. The proximity of the Sgr A complex (a bright radio source) meant that this structure could potentially appear in the sidelobes of our observations: producing artifacts in our Radio Arc NTF observations. Observations centered on Sgr A$\rm^*$ were made as a precaution to correct for any such effects found in our data. Fortunately, no such sidelobe contamination was detected in our observations of the Radio Arc NTFs.

\subsection{Calibration} \label{sec:calib}
\begin{figure}
    \centering
    \includegraphics[width=9cm]{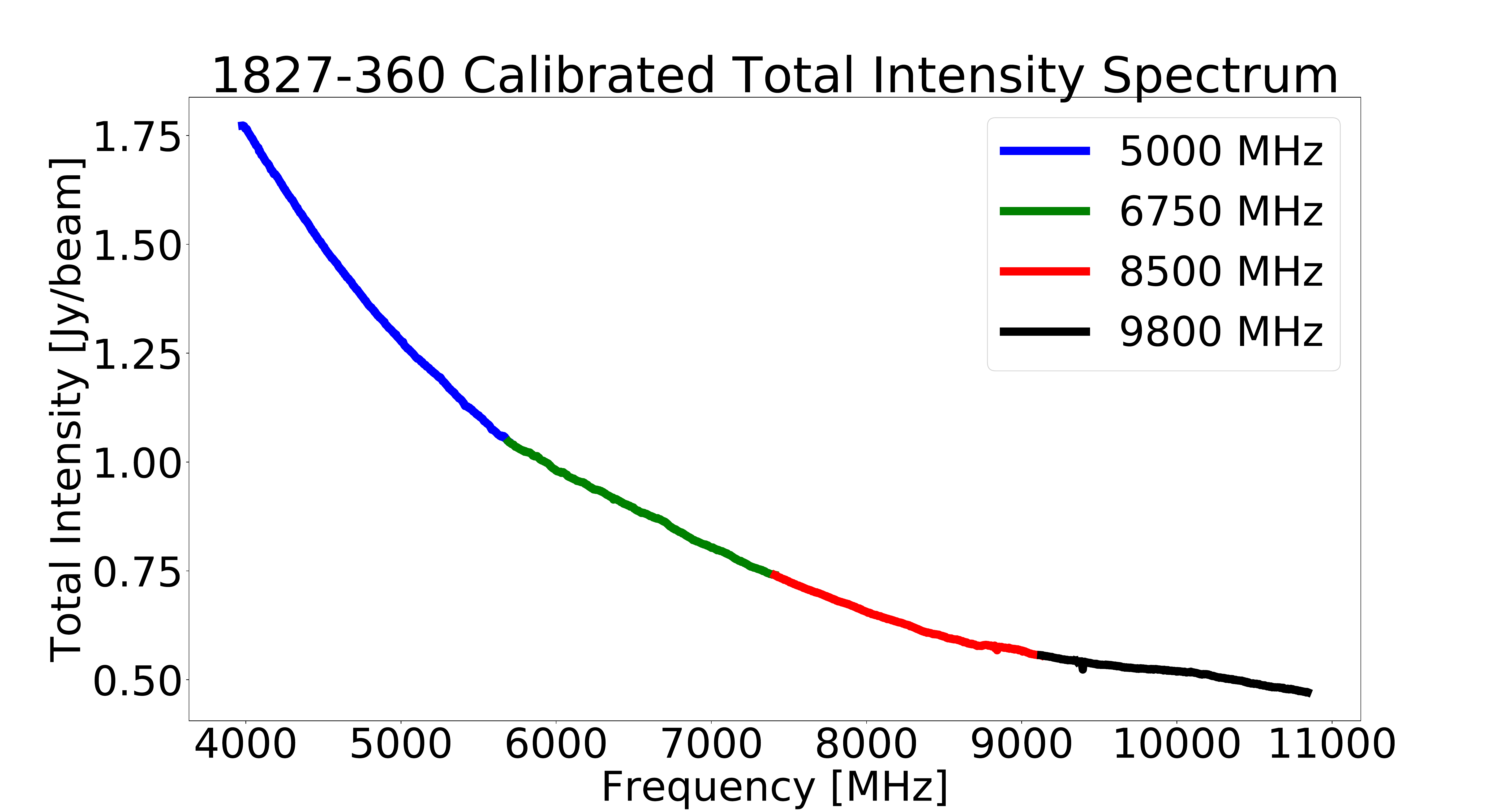}
    \caption{Calibrated total intensity spectrum of secondary calibrator PKS 1827-360 over the full frequency range of our observations. The different frequency bands used to make the spectrum are individually colored as marked in the legend in the upper right of the figure.}
    \label{fig:cal_spec}
\end{figure}
Three calibrators were observed at each observing frequency. PKS 1934-638 was used as the flux and bandpass calibrator, PKS 1827-360 was used as the polarization calibrator, and PKS 0823-500 was used as a backup calibrator. Our observations of the primary and secondary calibrators were of high enough quality that it was never necessary to work with our backup calibrator. However, the backup calibrator was used to confirm that the calibration procedures we used were correct.

We first flagged the data of the primary flux calibrator to remove any significant RFI. Additional flagging was employed for each individual day-frequency data block of the primary calibrator to account for issues like wind-stows; however, not many such custom flagging considerations were needed. This flagging was performed and applied using the task \textit{pgflag} in MIRIAD \citep{Sault1995}.

After flagging the primary calibrator data, antenna gain, delay, and passband calibrations were applied on each day-frequency block of data. In addition, phase and polarization calibration was performed on each day-frequency block. The calibration solutions were then applied to the primary and secondary calibrator. An additional iteration of phase and polarization calibration was performed on the secondary calibrator to improve the phase and polarization calibration solutions. The PKS 1827-360 calibration solutions were then applied to the observations of the Radio Arc NTFs. The calibration solutions for the calibrators  were found using the MIRIAD tasks \textit{mfcal} and \textit{gpcal}.

After one pass of these calibration steps it was necessary to perform another pass of calibrations using the solutions of the previous calibration runs. After this second pass, the phase solutions for the bandpass were flat across the day-frequency data, and so these final calibration solutions were applied to the Radio Arc NTF data.

The calibrated total intensity spectrum of PKS 1827-360 is shown in Figure \ref{fig:cal_spec}. Though there are some portions of the spectrum impacted by RFI, such as the region around 9400 MHz, the spectrum is quite continuous even across adjacent frequency bands. Only isolated channels from 4.0 $-$ 10.8 GHz are significantly effected by RFI. The continuity of the final calibrated spectrum indicates a high quality calibration of our observations.

\subsection{Imaging} \label{sec:imag}

Imaging of our observations was done using the Common Astronomy Software Applications (CASA) package after converting the MIRIAD uv data sets to CASA measurement sets \citep{McMullin2007}. This conversion was done using the CASA task \textit{importmiriad}. Multiple sets of images were produced using different imaging routines as described in detail below.

\subsubsection{Total Intensity Imaging} \label{sec:I_imag}
To produce total intensity images like the ones shown in Figures \ref{fig:ra_I_1} and \ref{fig:ra_I_2}, we used a Multi-Frequency Synthesis (MFS) cleaning routine that produced a single total intensity image for an entire spectral band of data. The cleaning procedure followed the method detailed in \citet{Clark1980} using a Briggs weighting of 0.5 and 100,000 iterations. This cleaning procedure was implemented in CASA using the task \textit{tclean}, resulting in 2D images of the total intensity distribution of the Radio Arc NTFs for each of the frequency bands observed. These images were mosaics of the individual pointings of the Radio Arc NTFs (two pointings at 2100 MHz, six pointings otherwise).

\subsubsection{Generating Cubes for RM Analysis} \label{sec:P_imag}
To study the depolarization and magnetic field properties of the Radio Arc NTFs, it is necessary to produce data cubes detailing the spectral dependence of the polarization. We produced cubes of the Stokes polarization products Q and U with axes of right ascension, declination, and frequency. To generate these images, we implemented a similar cleaning procedure to the one described in Section \ref{sec:I_imag}. However, instead of using an MFS cleaning, each channel in the data set was cleaned separately. In addition, far fewer cleaning iterations (100) were used per channel. Though the native ATCA spectral resolution of our data was 2 MHz, we averaged adjacent spectral channels together prior to cleaning to obtain channel widths of 4 MHz. This averaging increased the SNR of our polarized intensity within the resulting 4 MHz channels. We therefore had approximately 500 channels within a single frequency band. Each frequency band was cleaned independently and then combined into larger cubes consisting of the cleaned data of the four upper frequency bands listed in Table 1. These larger cubes consist of about 2000 channels.

With the Q and U cubes produced in this manner, we then generated cubes of linear polarization, $\rm{}P = \sqrt{Q^2+U^2}$ using the cleaned Q and U data. This operation was performed using the CASA task \textit{immath}.

Two sets of Q, U, and P cubes were produced in the manner described in the previous paragraphs. One set of cubes was made with the individual channels containing their native spatial resolutions, whereas the second set of cubes contain channels which had been uniformly smoothed to the largest beam size within the data. This largest beam size is 15.3$''$ x 6.4$''$.

\section{IMAGING RESULTS} \label{sec:res}

\subsection{Total Intensity Morphology} \label{sec:I_res}
\begin{figure*}
    \includegraphics[width=18cm]{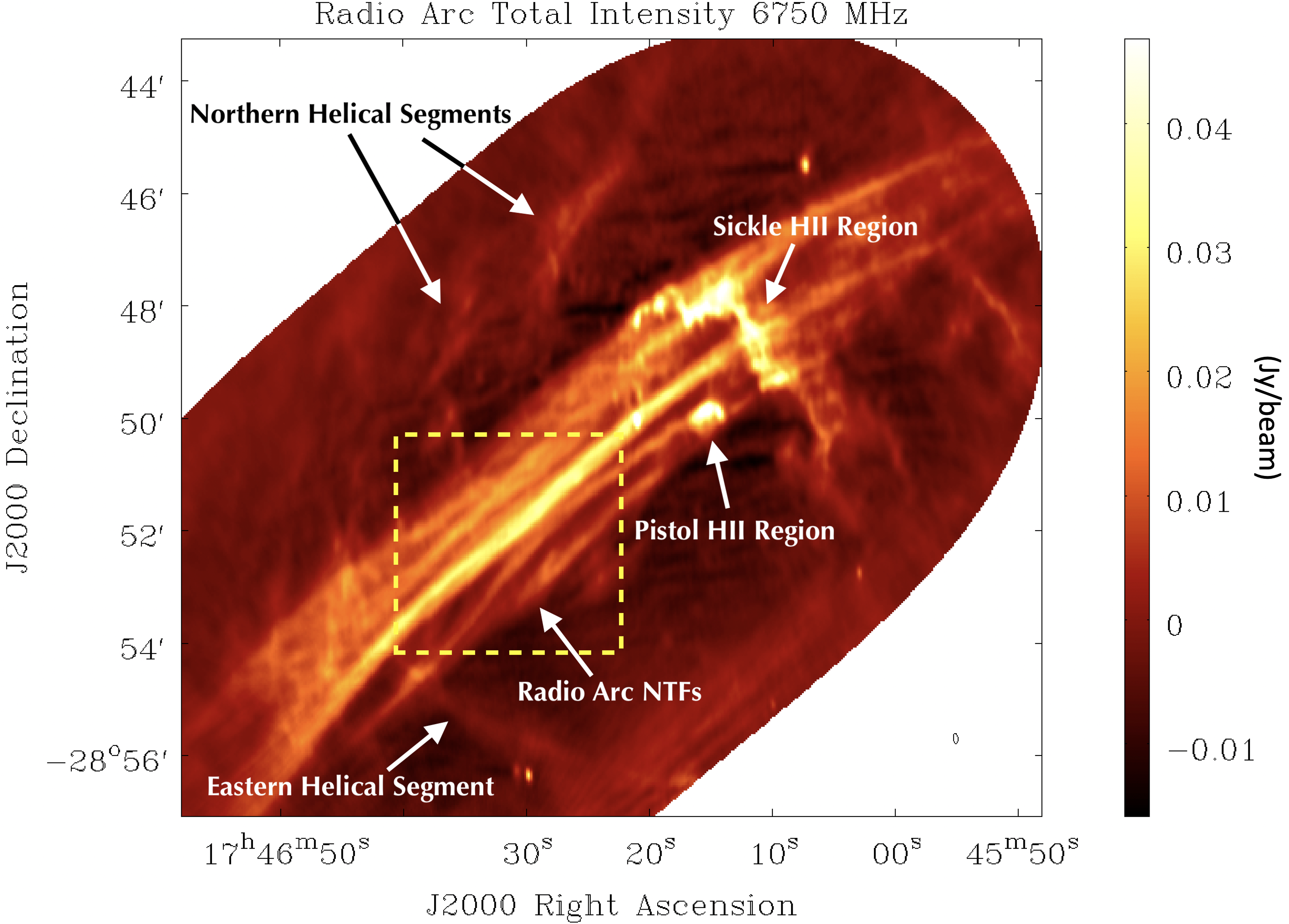}
    \caption{Representative total intensity distribution of the Radio Arc NTFs at 6750 MHz with a bandwidth of 2 GHz, a resolution of 10.9'' x 4.8,'' and an rms noise level of 7.3 mJy beam$\rm^{-1}$. Imaging was done over a 2.0 GHz frequency range from 5.7 to 7.7 GHz. The beam size is shown in the lower right corner of the figure. The rectangular region marked with dashed yellow lines indicates the field of view of the panels in Figure \ref{fig:ra_P} below. Notable features discussed in the text are marked and labeled.}
    \label{fig:ra_I_1}
\end{figure*}
\begin{figure*}
    \includegraphics[width=18cm]{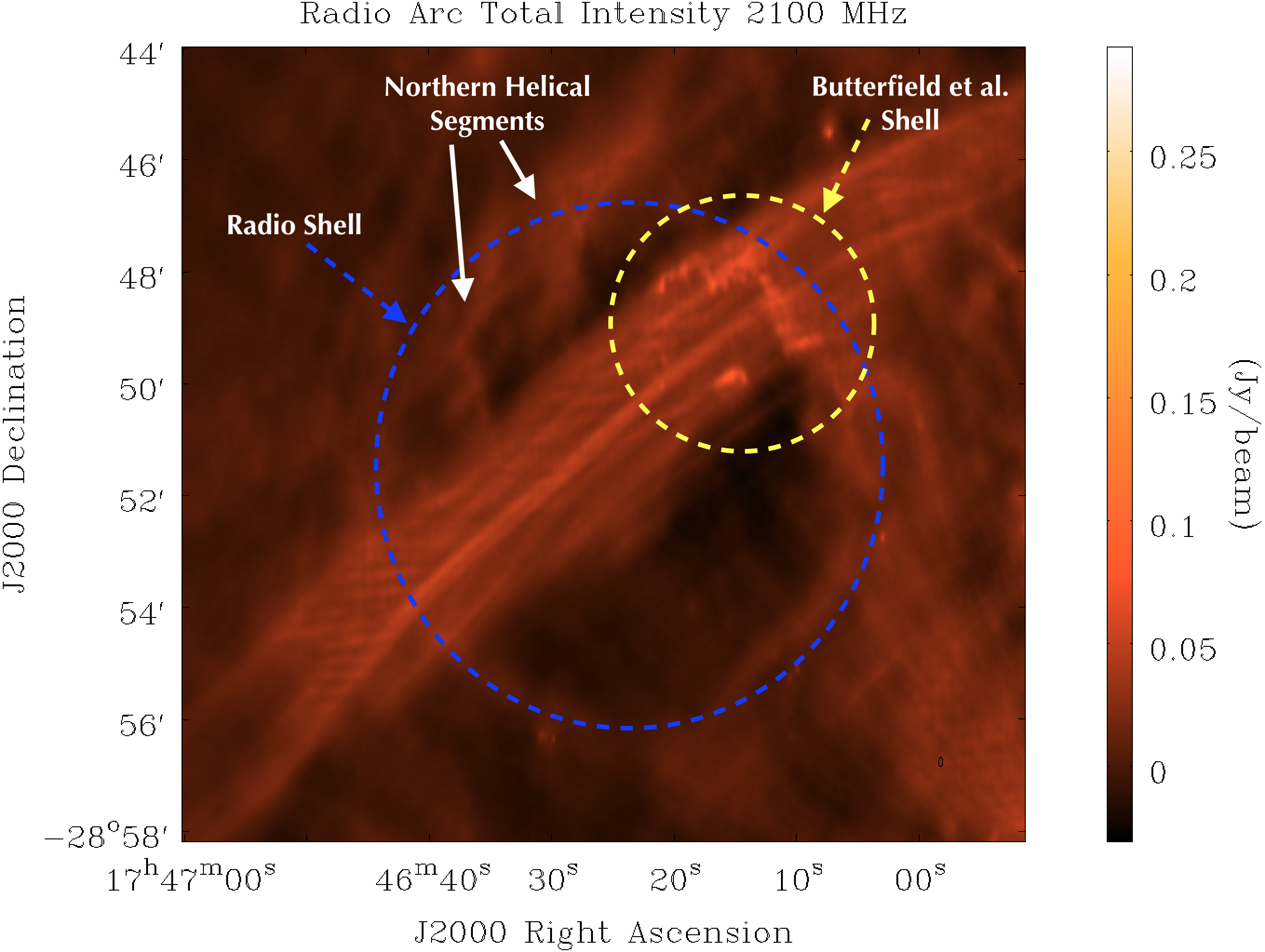}
    \caption{Total intensity distribution of the Radio Arc NTF region at 2100 MHz with a bandwidth of 2 GHz, a resolution of 9.4$''$ x 4.4$''$, and an rms noise level of 8.0 mJy beam$\rm^{-1}$. Imaging was done over a 2 GHz frequency range from 1.1 to 3.1 GHz. The beam size is shown in the lower right corner of the figure. The radio shell observed in the present study is marked with a blue dashed circle and a molecular shell seen in a previous study is marked with a yellow dashed circle \citep{Butterfield2018}. Notable features discussed in the text are marked and labeled with white arrows.}
    \label{fig:ra_I_2}
\end{figure*}
Figure \ref{fig:ra_I_1} shows a representative total intensity distribution for the Radio Arc NTFs at a central frequency of 6750 MHz. The region imaged reveals the Radio Arc NTFs, the `Sickle' and `Pistol' HII regions, and total intensity features labeled as helical segments  \citep{Pare2019,Lang1997,YM1987,Yusef-Zadeh1987a}. The term `helical segments' is derived from the interpretation of these structures in \citet{YM1987} and others where these total intensity features were thought to be segments of a larger helical structure encompassing the Radio Arc NTFs. 

In the southeastern portion of Figure \ref{fig:ra_I_1}, the NTFs become more faint and diffuse. This decrease in brightness -- also observed in \citet{Yusef-Zadeh1987a} and \citet{Pare2019} -- could be a result of limitations in the primary beams of the observations involved. Single-dish observations of the region, as well as observations made at higher and lower declinations, reveal a continuation of the NTFs beyond the extent of our observations \citep{Tsuboi1995}.

We also detect multiple helical segments which are superimposed on or near the Radio Arc NTFs and which possibly pass through the NTF system (Figure \ref{fig:ra_I_1}). These features are more clearly observed in Figure \ref{fig:ra_I_2}. The helical segments detected to the North of the NTF system have been reported previously \citep{Yusef-Zadeh1987a,YM1987,Inoue1989,Pare2019}, and are possibly part of a larger structure. The easternmost helical segment is seen to cross the NTF system, appearing both above and below the Radio Arc NTFs.

The helical segments seem to be superimposed on a shell of radio emission, as seen and marked with a blue circle in Figure \ref{fig:ra_I_2}. This shell encompasses the observed portion of the Radio Arc NTFs. \citet{Pare2019} posited that the helical segments and this shell are components of a single large structure, and the images presented here are consistent with this hypothesis. In this scenario, the helical segments could be density enhancements of the shell, with the more diffuse portions of the shell being undetected in the present observations.

This same shell (marked in blue in Figure \ref{fig:ra_I_2}) was also observed at infrared wavelengths by \citet{Simpson2007}. A smaller molecular shell was also identified within this same region by \citet{Butterfield2018} as shown by the yellow circle in Figure \ref{fig:ra_I_2}. This smaller shell is not visible in our data because its presence is manifested by its molecular spectral lines rather than radio continuum emission \citep{Butterfield2018}. The presence of both of these shells in this region indicates the complexity of the environment local to the Radio Arc NTFs.

One reason we suspect the larger contiguous shell structure may not be fully detected in our ATCA observations is due to the limit in the largest angular size of structures we are able to resolve with our interferometric observations of the Radio Arc NTFs. Due to the lack of data at the origin of the uv-plane for interferometer observations (so called `zero-spacings'), large-scale structures are undetected. The shell of radio emission has a diameter (as seen in our observations) of 570'' ($\rm\sim{}$22 pc assuming a distance of 8.0 kpc to the GC). At the central frequency of our lowest frequency band (5000 MHz), fully resolving the shell would require a shortest baseline length of $\rm\sim{}$13 m. In our observations, however, the shortest baseline length is $\rm\sim{}$46 m, which corresponds to a largest sampled angular size of $\rm\sim$150''. As a result, structures larger than this angular size would be undetected in our observations.

\subsection{Polarized Intensity Morphology} \label{sec:pol_morph}
\begin{figure*}
    \centering
    \includegraphics[width=18cm]{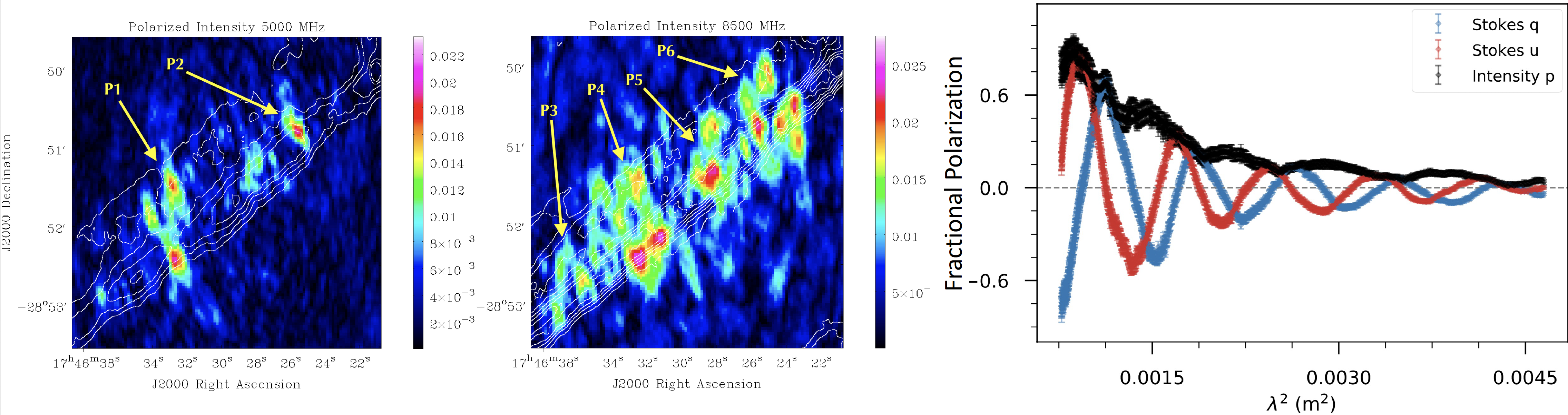}
    \caption{Polarized intensity emission from the region marked in Figure \ref{fig:ra_I_1}  with total intensity distributions overlayed as contours at 5000 MHz (left panel) with a resolution of 11.5'' x 4.8'' and contour levels of 0.1, 0.15, 0.2, 0.25, and 0.3 x the peak total intensity value of 0.091 mJy beam$\rm^{-1}$ and 8500 MHz (middle panel) with a resolution of 7.1'' x 4.2'' and contour levels of 0.05, 0.1, 0.15, 0.2, 0.25, and 0.3 x the peak total intensity value of 0.0641 mJy beam$\rm^{-1}$. Extended polarized features discussed in the text are marked with yellow arrows. The units of the color bars are Jy/beam. The right panel shows the fractional q, u, and p spectra for a representative line-of-sight.}
    \label{fig:ra_P}
\end{figure*}
Figure \ref{fig:ra_P} displays representative polarized intensity distributions obtained from the ATCA data. The left panel shows the polarized intensity seen at 5000 MHz, with total intensity contours at the same frequency overlayed. The middle panel reveals the same information for a channel at a frequency of 8500 MHz. The right panel shows example spectra of the Stokes Q, U, and polarized intensity for a single pixel. The polarized intensity exhibits a characteristic ``clumpiness'' where regions of polarized intensity are separated by regions of almost complete depolarization. The discontinuous nature of the polarized intensity is most prominent at lower frequencies, as can be seen by comparing the distributions shown in the left and middle panels of Figure \ref{fig:ra_P}. The polarized intensity becomes more continuous at higher frequencies, which agrees with previous results of the Radio Arc NTFs \citep{Pare2019}.

The polarized intensity is concentrated within a narrow region of the Radio Arc NTFs from RA = 17h 46m 22s to RA = 17h 46m 38s. This region roughly corresponds with the dashed yellow box in Figure \ref{fig:ra_I_1}. The thermal emission from the large electron densities in the HII regions seen in the upper right of Figure \ref{fig:ra_I_1} could be depolarizing any polarized emission that might be present in the Radio Arc NTFs in this region, thus explaining the decrease in polarized intensity in that region. However, it is unclear why the polarization does not extend into the region of the Radio Arc NTFs seen in the lower left portion of Figure \ref{fig:ra_I_1} given that other NTFs are observed to be polarized throughout their extents \citep{Gray1995,Lang1999a,Lang1999b}.

We detect extended polarized intensity structures that exist in regions of low total intensity. These structures have been marked in Figure \ref{fig:ra_P}. There are two of these features in the left panel, labeled P1 and P2, and four of these features in the middle panel, labeled P3 - P6. P1 and P4 occur at the same spatial position within the Radio Arc NTFs as do P2 and P6. This spatial correspondence indicates that these could be the same structures which are being seen at multiple frequencies. Previous studies of the Radio Arc NTFs have detected these polarized extensions in the Radio Arc NTFs as well (e.g. \citet{Inoue1989,Pare2019}). These structures remain visible even after the polarized intensity is de-biased at the 3$\rm\sigma$ level, indicating that these structures are well above the rms noise of the polarized intensity.

These extended polarized intensity features are found to have large fractional polarizations greater than 0.7. One explanation for this could be that not all of the total intensity emission associated with these features is being detected in our observations. This could occur if the total intensity is associated with a structure more extended than the largest angular scales that our observations are sensitive to (Section \ref{sec:I_res}). Such an effect would disproportionately impact the total intensity, since the total intensity structure is continuous whereas the polarized intensity is comprised of smaller, discrete polarized clumps due to depolarization effects. This could indicate that these polarized intensity structures are spatially delimited portions of a larger structure local to the Radio Arc NTFs.

The right-hand panel of Figure \ref{fig:ra_P} reveals that the polarized intensity spectrum does not exhibit a uniform trend with wavelength, but contains small oscillations. Similar divergences from a linear trend are seen in the total intensity spectra obtained for the Radio Arc NTFs. These unusual spectral features are likely not the result of a problematic calibration, because we saw in Section \ref{sec:calib} in Figure \ref{fig:cal_spec} that the spectrum obtained for the calibrator was smooth and contiguous. Rather, the complicated spectral features for the Radio Arc NTFs are likely a result of the complicated nature of the Radio Arc NTFs and environment along the line-of-sight.

\subsection{Rotation Measure Characteristics} \label{sec:RM_char}
\begin{figure}
    \centering
    \includegraphics[width=8.5cm]{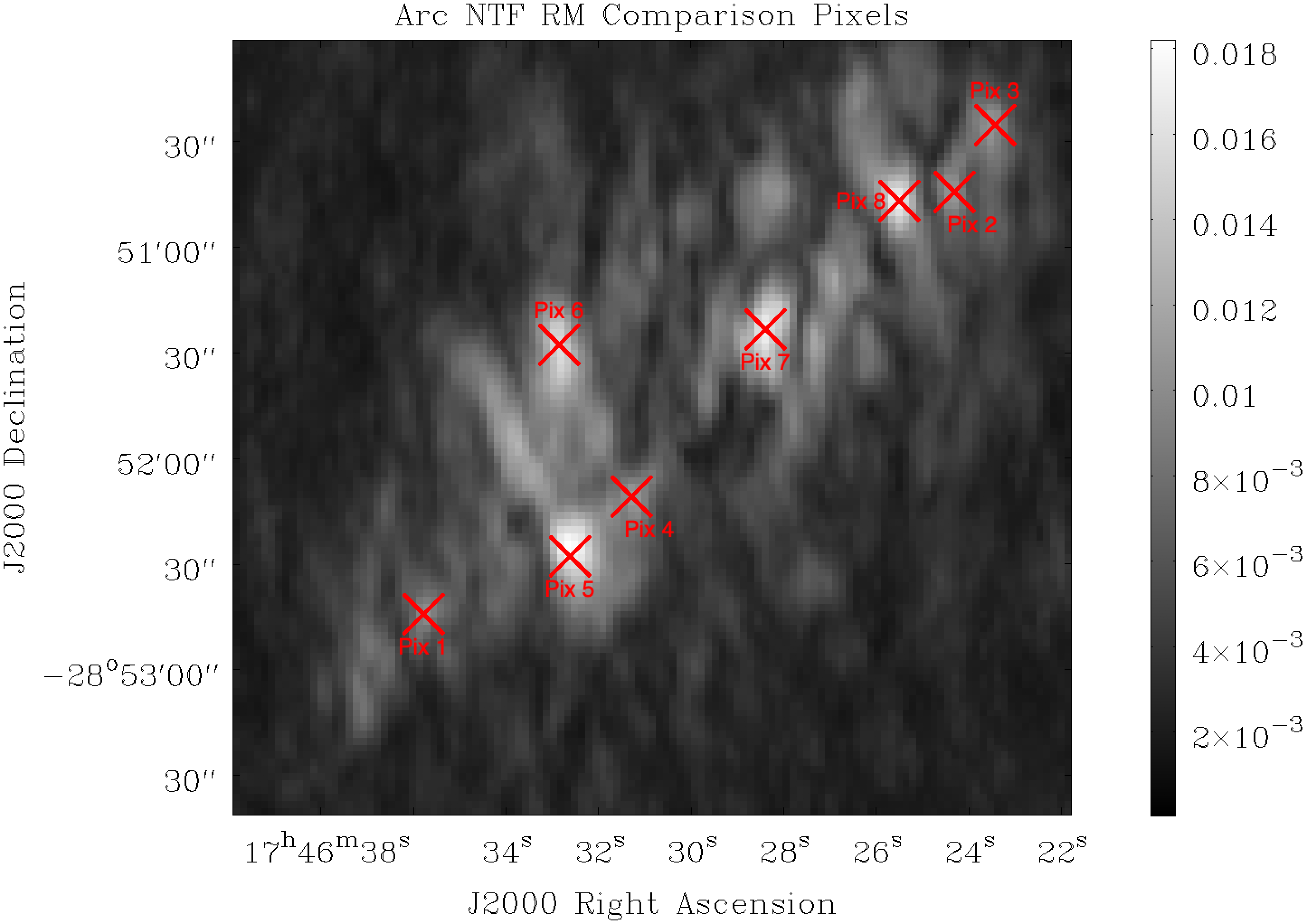}
    \includegraphics[width=8.5cm]{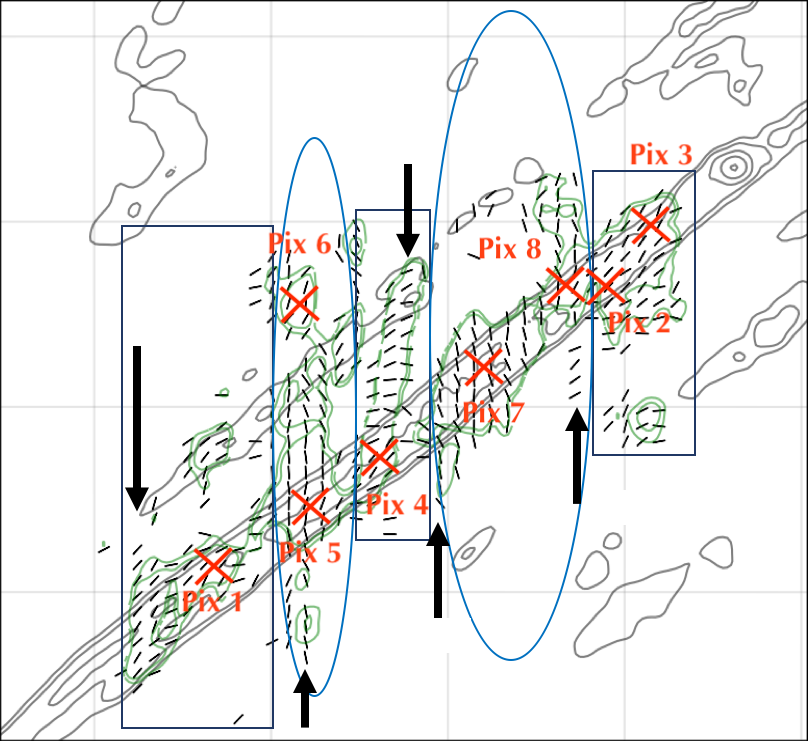}
    \caption{\textbf{Upper panel:} Locations of the pixels used to compare RM values are shown as red crosses overlayed on a grayscale map of the polarized intensity at a central frequency of 7500 MHz. The color bar is in units of Jy/beam. \textbf{Lower panel:} figure 8 of \citet{Pare2019} showing the magnetic field distribution with total (black) and polarized (green) intensity contours. Black rectangles mark regions of parallel magnetic field and blue ellipses mark regions of rotated magnetic field. Locations of the pixels used for the RM analysis in the present work are shown with red crosses with `Pix' numbering abbreviations and the locations of the extended polarized intensity features discussed in Section \ref{sec:P_imag} are shown with black arrows.}
    \label{fig:pix_loc}
\end{figure}
The polarized intensity of the Radio Arc NTFs allows us to analyze how the observed polarization angle changes from its initial value due to Faraday effects produced by magneto-ionic media along the line-of-sight. Different regions of polarized emission within a single resolution element will experience different amounts of Faraday rotation. In addition, multiple rotating media along the line-of-sight will rotate the polarized emission in distinct ways. To fully describe the rotation of a particular line-of-sight accounting for multiple rotating media it is convenient to describe the rotation using a parameter known as Faraday depth \citep{Burn1966}:
\begin{equation}
    \rm \phi = \frac{e^3}{2\pi{}m_e^2c^4}\int_0^L\!n_eB_{||}\,dl \label{eq:RM}
\end{equation}
where $\rm\phi$ is the Faraday depth in the region, e and $\rm{}m_e$ is the charge and mass of an electron, c is the speed of light, $\rm{}n_e$ is the electron number density, $\rm{}B_{||}$ is the strength of the line-of-sight component magnetic field within the region, L is the distance between the source and the observer, and dl is the deferential distance element along the line-of-sight. We follow the convention of \citet{Brentjens2005} by defining a Faraday thin source as one where $\rm\lambda^2\Delta\phi << 1$ and a Faraday thick source as one where $\rm\lambda^2\Delta\phi >> 1$ where $\rm\Delta\phi$ is the extent of the medium in Faraday depth space and $\rm\lambda$ is the wavelength.

\begin{deluxetable}{|c|c|c|c|c|}[ht!]
\tablecaption{Comparison of RM Values}
\tablecolumns{5}
\tablenum{2}
\tablewidth{0pt}
\tablehead{
\colhead{Pix} & \colhead{RA} & \colhead{DEC} & \colhead{$\rm\phi$ [rad m$\rm^{-2}$]} & \colhead{RM [rad m$\rm^{-2}$]\tablenotemark{a}}
}
\startdata
1 & 17:46:36 & -28.52.44 & -4000 & -4000 \\
2 & 17:46:24 & -28.50.44 & -4000 & -3800 \\
3 & 17:46:23 & -28.50.25 & -3900 & -3900 \\
4 & 17:46:31 & -28.52.11 & -2900 & -2900 \\
5 & 17:46:33 & -28.52.28 & -3100 & -3100 \\
6 & 17:46:33 & -28.51.28 & -2600 & -2700 \\
7 & 17:46:28 & -28.51.24 & -1100 & -1600 \\
8 & 17:46:26 & -28.50.46 & -2400 & -2400 
\enddata
\tablecomments{Pix indicates the pixel number as shown in Figure \ref{fig:pix_loc}, RA indicates the right ascension location of each pixel expressed in HH:MM:SS, DEC shows the declination of each pixels expressed as DD.MM.SS, RM Synth indicates the peak Faraday depth value obtained for the given pixel using our ATCA data, and 2019 RM indicates the RM value obtained for the given line-of-sight as seen in \citet{Pare2019}.}
\tablenotetext{a}{Values taken from \citep{Pare2019}}
\label{table:RM_compare}
\end{deluxetable}

In the simple case where there is a source of polarized emission and rotation due only to a foreground magneto-ionic medium, the Faraday depth is equal to the rotation measure (RM) and can be measured by analyzing how the polarization angle changes as a function of wavelength squared:
\begin{equation}
    \rm \chi = RM*\lambda^2+\chi_0 \label{eq:RM_LS}
\end{equation}
where $\rm\chi$ is the observed polarization angle [rad], $\rm\lambda^2$ is the wavelength squared of the observation [$\rm{}m^2$], RM is the rotation measure of the foreground medium [rad $\rm{}m^{-2}$]. and $\rm\chi_0$ is the intrinsic polarization angle of the source [rad]. In these situations, the RM of a line-of-sight can be extracted by fitting a linear model to the polarization angle as a function of wavelength squared.

This linear fitting method has been used to determine the RM values of several NTFs \citep{Gray1995,Lang1999a,Lang1999b,Pare2019}. NTF RM magnitudes found using this method have been in the thousands of rad m$\rm^{-2}$. This method, however, assumes the rotation along the line-of-sight is characterized by only a single foreground medium. Since the Radio Arc NTFs are embedded within such a complex environment within the GC, this assumption may not be applicable.

For this work, we use the techniques of RM-synthesis and QU-fitting to robustly determine the RM due to multiple components along the line-of-sight, including components internal to the source of the polarized emission \citep{Brentjens2005,OSullivan2012}. To test the quality of our ATCA data and the validity of the RM-Synthesis method, we compare the peak Faraday depth value found through RM-synthesis with previous RM values for the Radio Arc NTFs found using the linear fitting method. 

\citet{Pare2019} defined parallel and rotated magnetic field regions for the Radio Arc NTFs, in each of which the magnetic field maintains a roughly constant orientation within a 10 degree range (see Figure \ref{fig:pix_loc}). Regions of parallel magnetic field are marked with rectangular zones and regions of rotated magnetic field are marked with elliptical zones. We ran RM-synthesis over 8 pixels that are in locations of peak polarized intensity and varying magnetic field orientation as shown in Figure \ref{fig:pix_loc}. Pixels 1 - 4 are located in regions of parallel magnetic field whereas pixels 5 - 8 are located in regions of rotated magnetic field.

The RMtools package (Purcell et al. in prep) was used to perform RM-Synthesis on the 8 target pixels. We extracted the peak Faraday depth values obtained for these pixels and compared them with the published RM values found in \citet{Pare2019} in Table 2. There is general agreement for 7 out of 8 of the pixels. The pixel with different values for the different methods corresponds with a region of comparatively low RM magnitude. This raises the possibility that the two methods generally give discrepant RM values at locations of lower RM magnitude. We conclude that we see generally the same RM structure in our ATCA data as is seen in the 2019 VLA data set of \citet{Pare2019}

\subsection{Predictions of Best-Fitting Models}
The structure in the polarized intensity and RM, as seen in Sections \ref{sec:pol_morph} and \ref{sec:RM_char}, indicates that complex processes could be involved in producing the polarization and RM we observe. By simultaneously modeling how the linear polarization products, Stokes Q and U, the polarized intensity, and the polarization angle vary as a function of frequency, it is possible to test different hypotheses regarding the origin of the RM toward the Radio Arc NTFs by modeling depolarization mechanisms and comparing the models to the data. This modeling is known as QU-fitting, and has been used to analyze a number of cosmic sources \citep{OSullivan2012,OSullivan2017,Kaczmarek2018}.

This modeling will allow us to verify the presence of the intervening shell proposed by \citet{Pare2019}. In that paper, they proposed that density enhancements within this intervening shell were producing the regions of rotated magnetic field seen in their VLA data. By comparing the data to different models through this method, we can determine which hypothesis for the origin of rotation is best supported by the data. Through fitting the Stokes Q and U spectra, we would expect to obtain a different best-fit result in regions corresponding to the rotated magnetic field than in regions corresponding to the parallel magnetic field. 

\begin{figure}
    \centering
    \includegraphics[width=8.5cm]{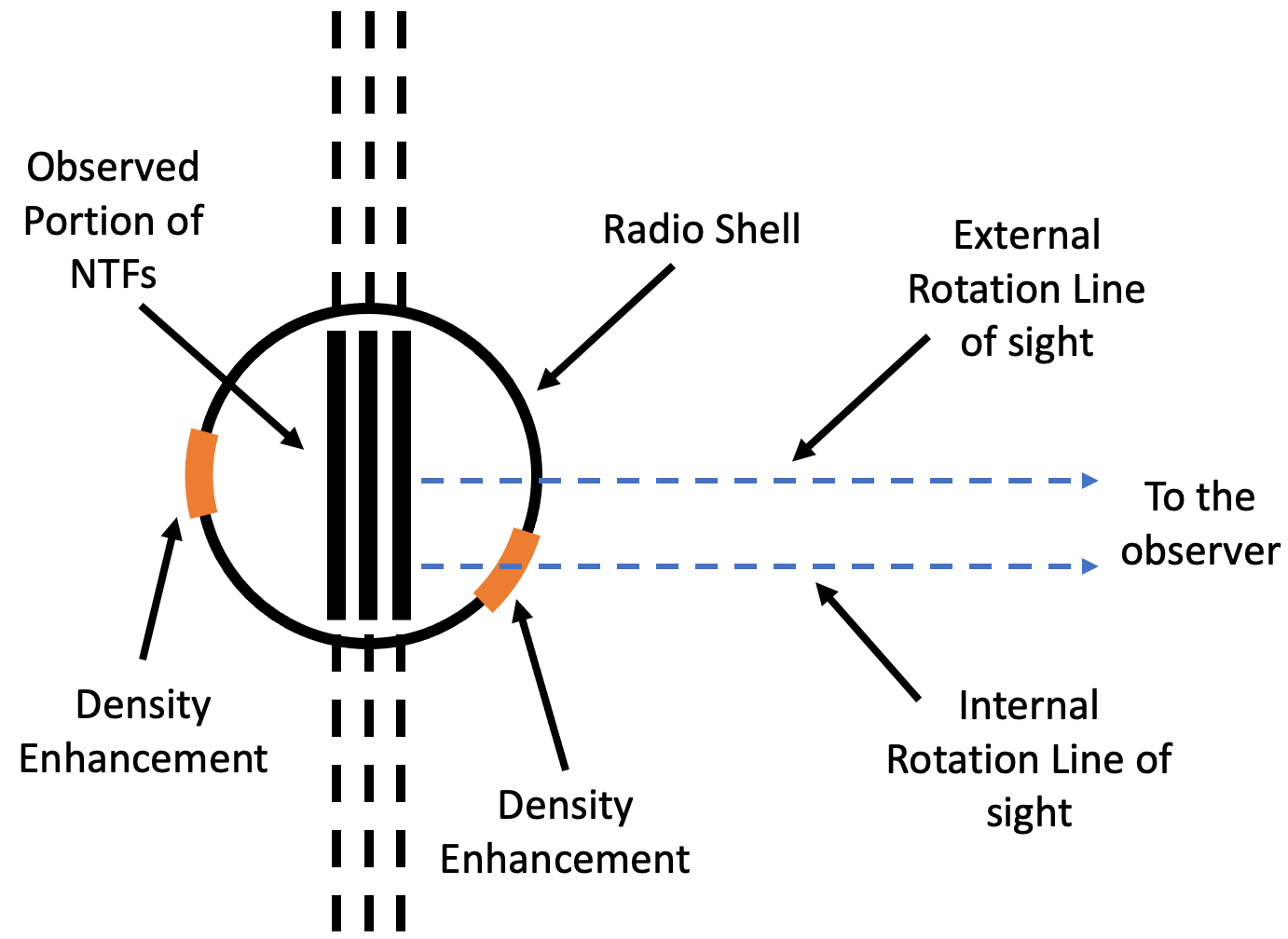}
    \caption{A sketch showing the possible physical environment local to the Radio Arc NTFs. The NTFs are shown as black parallel lines with the solid portion representing the observed portion of the NTFs and the dashed portions indicating the extensions of the NTFs beyound our FOV. The black circle represents the Radio Shell with the orange wedges indicating density enhancements within the shell. The dashed blue lines represent two different lines of sight to the observer, one which passes through a density enhancement of the shell, and one which does not.}
    \label{fig:mod_pic}
\end{figure}

For regions unaffected by the Radio Shell, we expect physical scenarios of the form shown in the upper line-of-sight in Figure \ref{fig:mod_pic} to be applicable, where the depolarization and rotation are caused by only external media located elsewhere in the GC and in the Galactic Disk. For regions of the Radio Arc NTFs affected by the Radio Shell, we would expect a physical scenario of the form represented by the lower line-of-sight in Figure \ref{fig:mod_pic} to apply. In this scenario, the Radio Shell is both an emitting and rotating structure, and so produces its own rotation and depolarization, leading to internal Faraday effects possibly being significant for such lines of sight. QU-fitting will allow us to verify whether these scenarios accurately reflect reality, thereby allowing us to assess whether the Radio Shell is responsible for the rotated magnetic field of the Radio Arc NTFs.

\subsection{Models Characterizing Internal and External Media}
To probe whether the Radio Shell is impacting the observed properties of the Radio Arc NTFs, we test different models of mechanisms which cause depolarization and rotation in cosmic sources. These mechanisms were chosen because they are processes commonly discussed and utilized in the literature \citep{Burn1966,Sokoloff1998,OSullivan2012,OSullivan2017,Kaczmarek2018}. The equations detailed below are derived from \citet{Sokoloff1998}.

\subsubsection{External Faraday Dispersion (EFD)}
EFD is produced when a foreground Faraday screen that contains a spatially varying magnetic field or electron density lies between the source of the emission and the observer. The rapid spatial variation of the magnetic field causes different lines of sight through the foreground medium to experience different amounts of Faraday rotation:

\begin{equation}
    \rm P = p_0e^{2i(\chi_0+RM\lambda^2)}e^{-2\sigma_{RM}^2\lambda^4} \label{eq:EFD}
\end{equation}
where P is the complex polarized intensity observed by the observer [Jy beam$\rm^{-1}$], $\rm{}p_0$ is the polarized intensity emitted by the source [Jy beam$\rm^{-1}$], RM is the rotation caused by the foreground medium [rad m$\rm^{-2}$], $\rm\lambda$ is the wavelength of the observation [m], $\rm\chi_0$ is the intrinsic polarization angle of the source [rad], and $\rm\sigma_{RM}$ is the standard deviation of the RM of the medium within the observing beam [rad m$^{-2}$]. 

\subsubsection{Differential Faraday Rotation (DFR)}
DFR is produced when a polarized source rotates its own emission. This internal rotation could be produced when the source medium has a non-negligible electron density and is in the presence of a significant line-of-sight component of the magnetic field. The DFR mechanism characterizes sources of this type that are in the presence of a uniform magnetic field:
\begin{equation}
\rm P = p_0\frac{\sin\left(2RM\lambda^2\right)}{2RM\lambda^2}e^{2i(\chi_0 + RM\lambda^2)} \label{eq:DFR}    
\end{equation}

\subsubsection{Internal Faraday Dispersion (IFD)}
IFD is produced in a similar manner to DFR, with the difference being that the magnetic field local to the source is randomly oriented and anisotropic:
\begin{equation}
    \rm P = p_0e^{2i\chi_0}\frac{1 -e^{4iRM\lambda^2-2\sigma_{RM}^2\lambda^4}}{2\sigma_{RM}^2\lambda^4-4iRM\lambda^2} \label{eq:IFD}
\end{equation}
Each of the three models above describes a single medium along the line-of-sight, but because of the complexity of the GC region multiple media will likely be needed to accurately model the data. To do so, we sum individual media components such that:
\begin{equation}
    P = P_1 + P_2 + P_3 + ... + P_N \label{eq:P_sum}
\end{equation}
where P is the final polarization as observed at the telescope and each of the component parts (like $\rm{}P_1$) is the polarization characterized by a single medium using one of Equations \ref{eq:EFD} - \ref{eq:IFD} above.

\section{MODEL FITTING RESULTS} \label{sec:model_res}
\begin{figure*}[hb!]
    \gridline{\fig{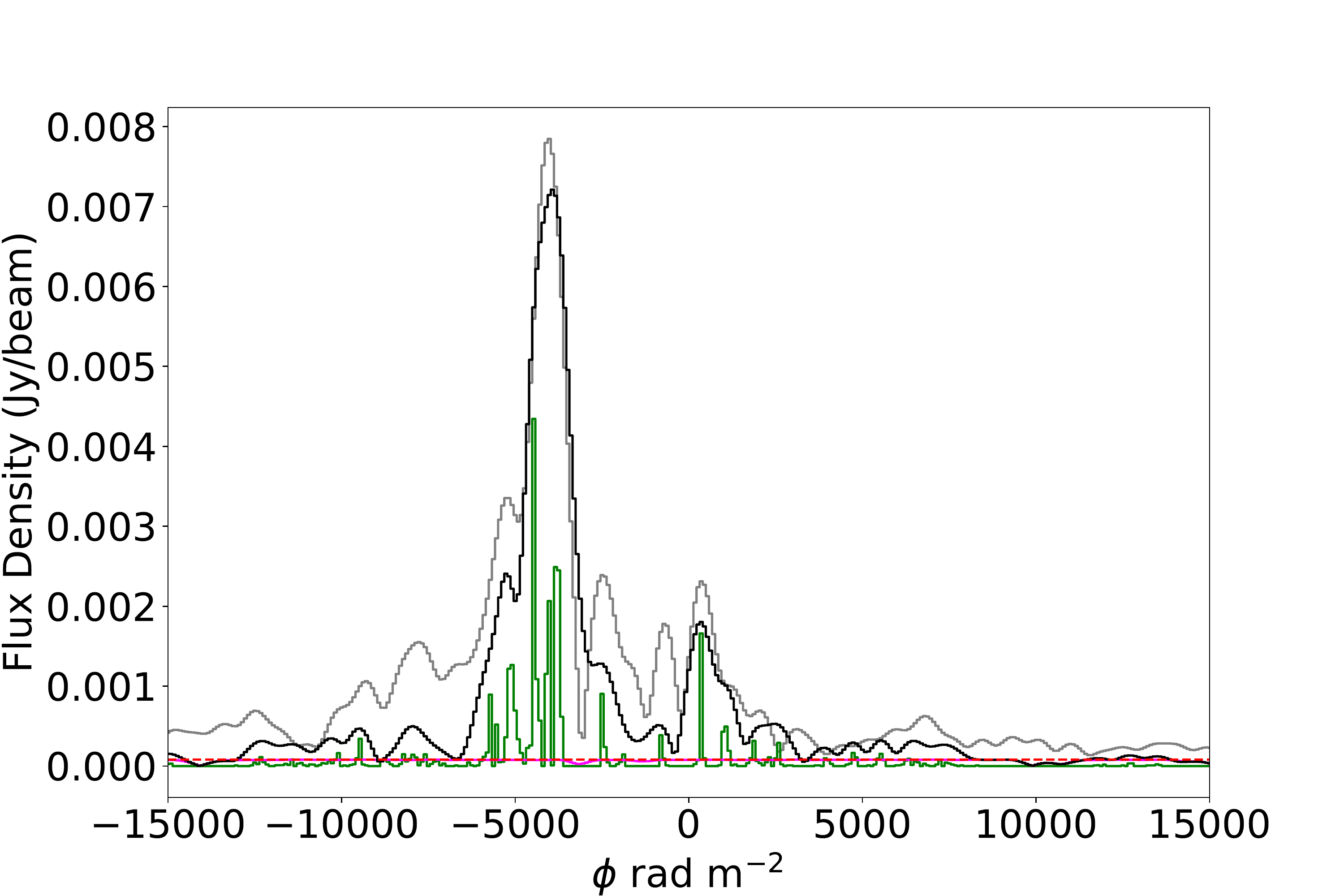}{0.40\textwidth}{1}
    \fig{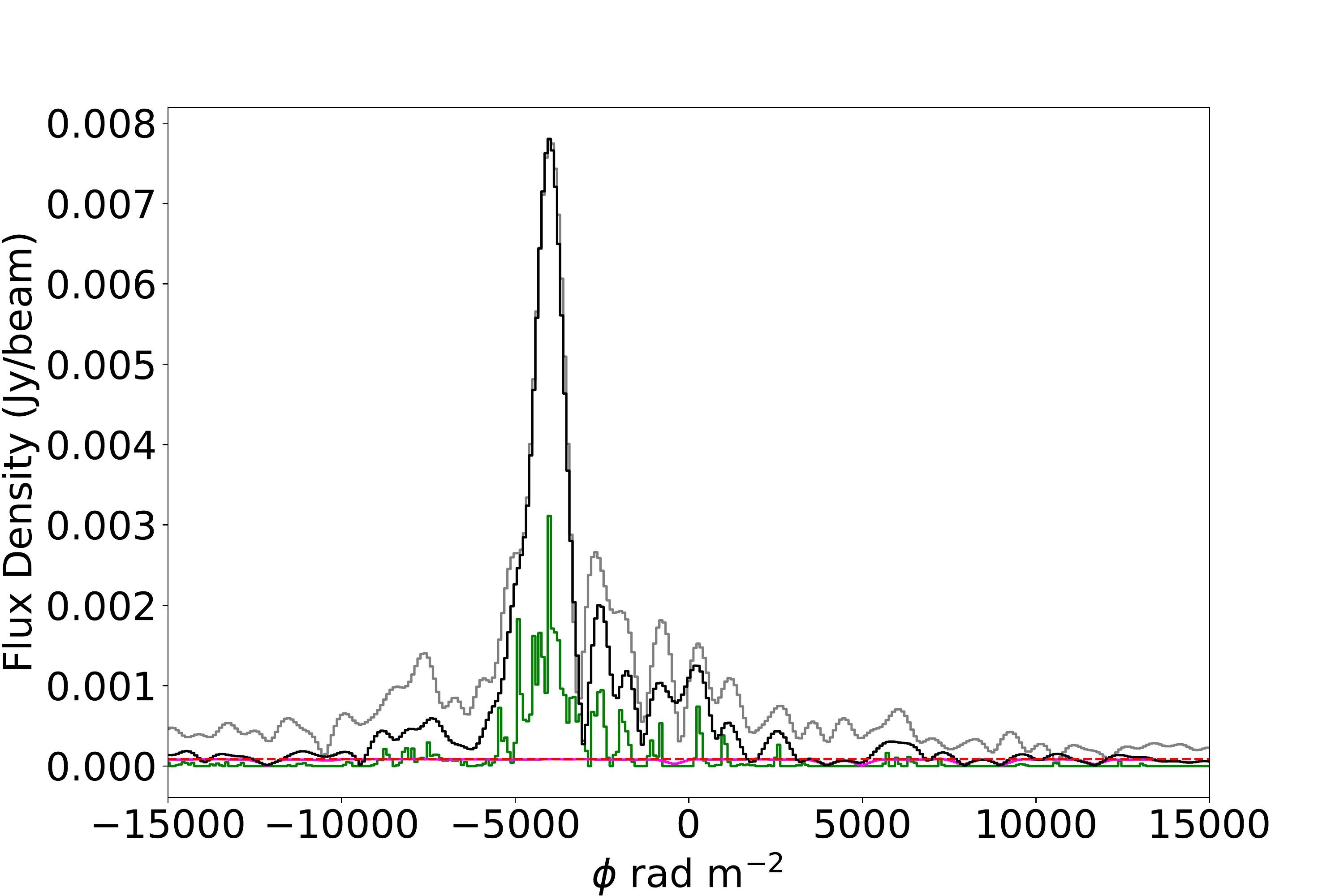}{0.40\textwidth}{2}
    }
    \gridline{\fig{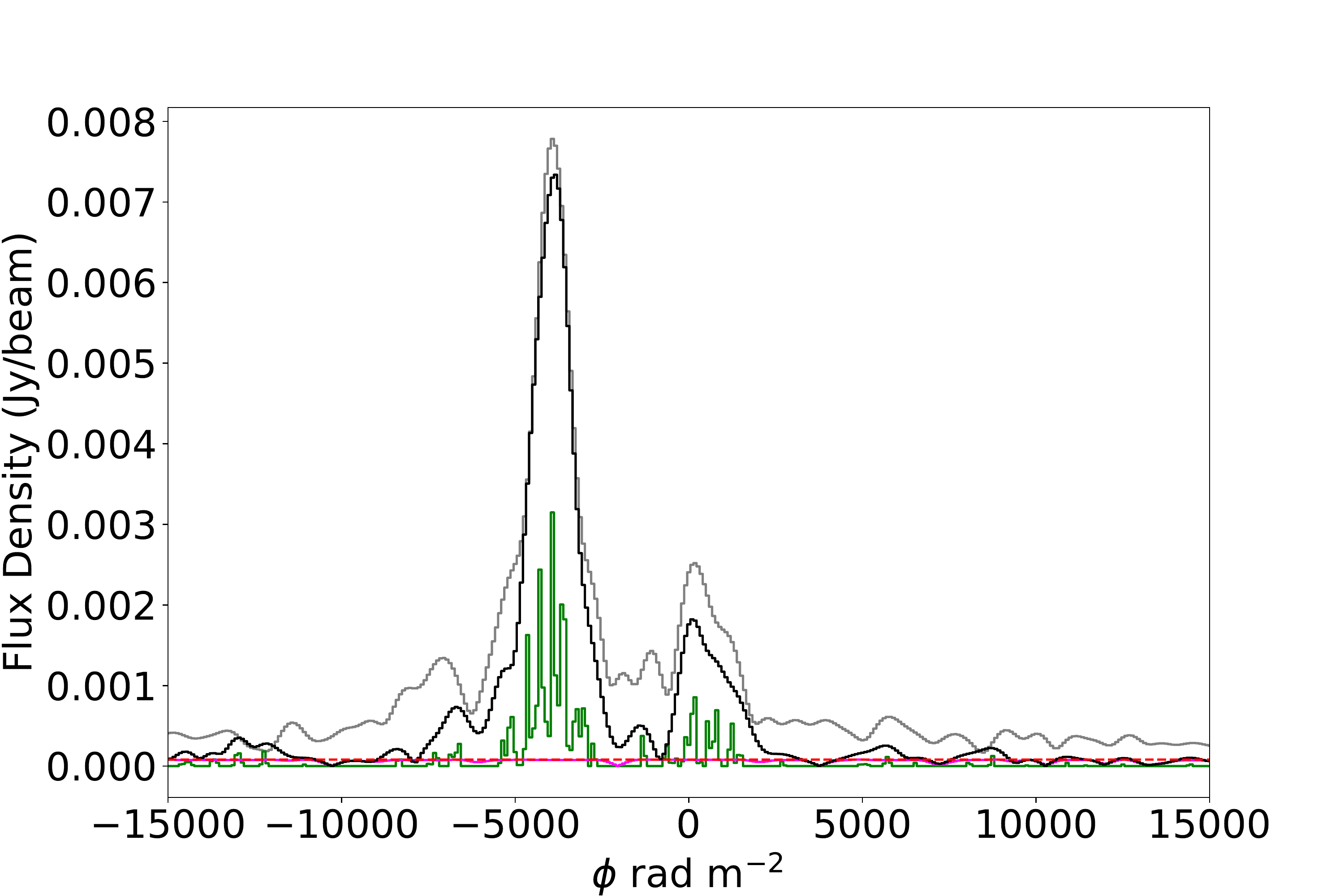}{0.40\textwidth}{3}
    \fig{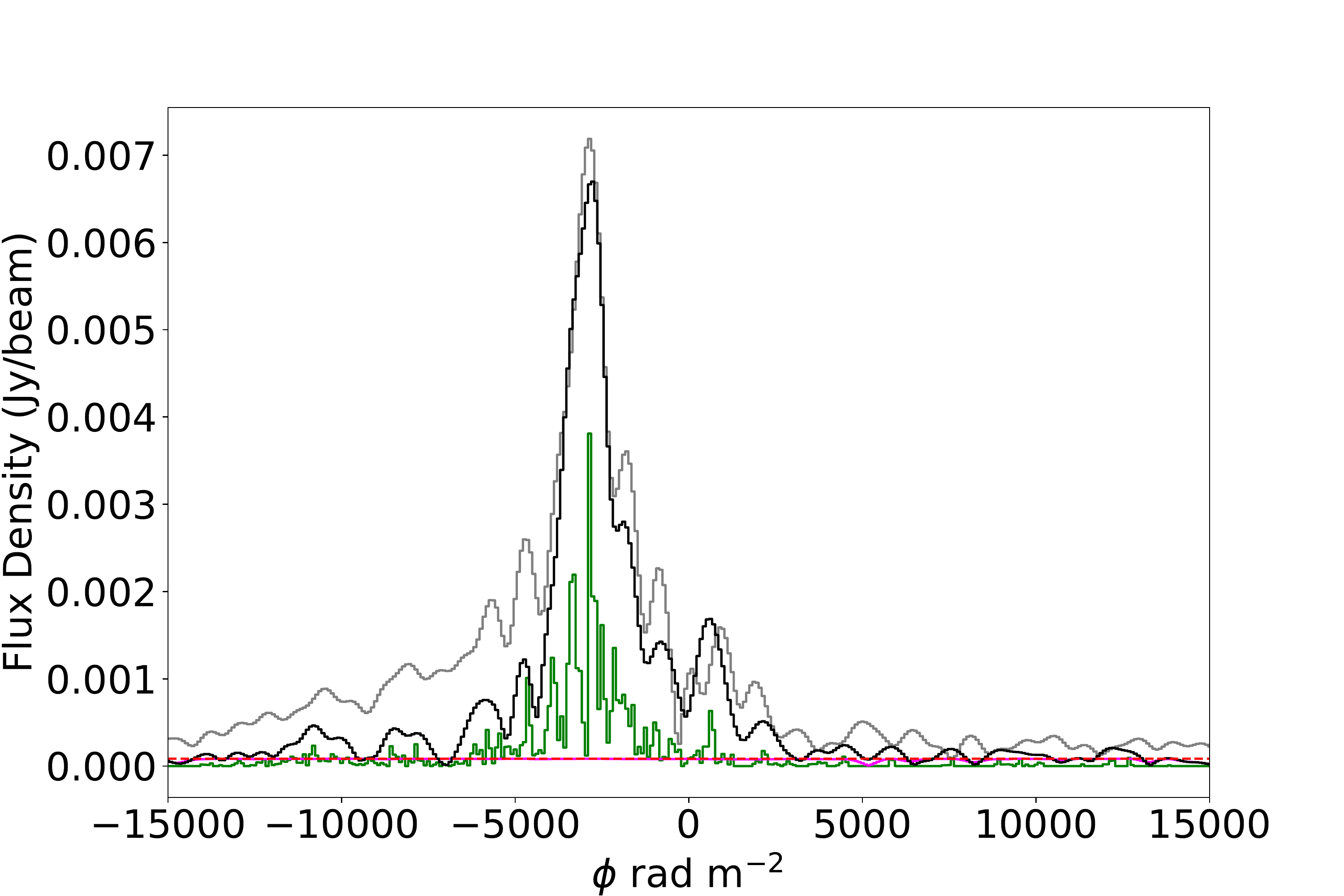}{0.40\textwidth}{4}
    }
    \gridline{\fig{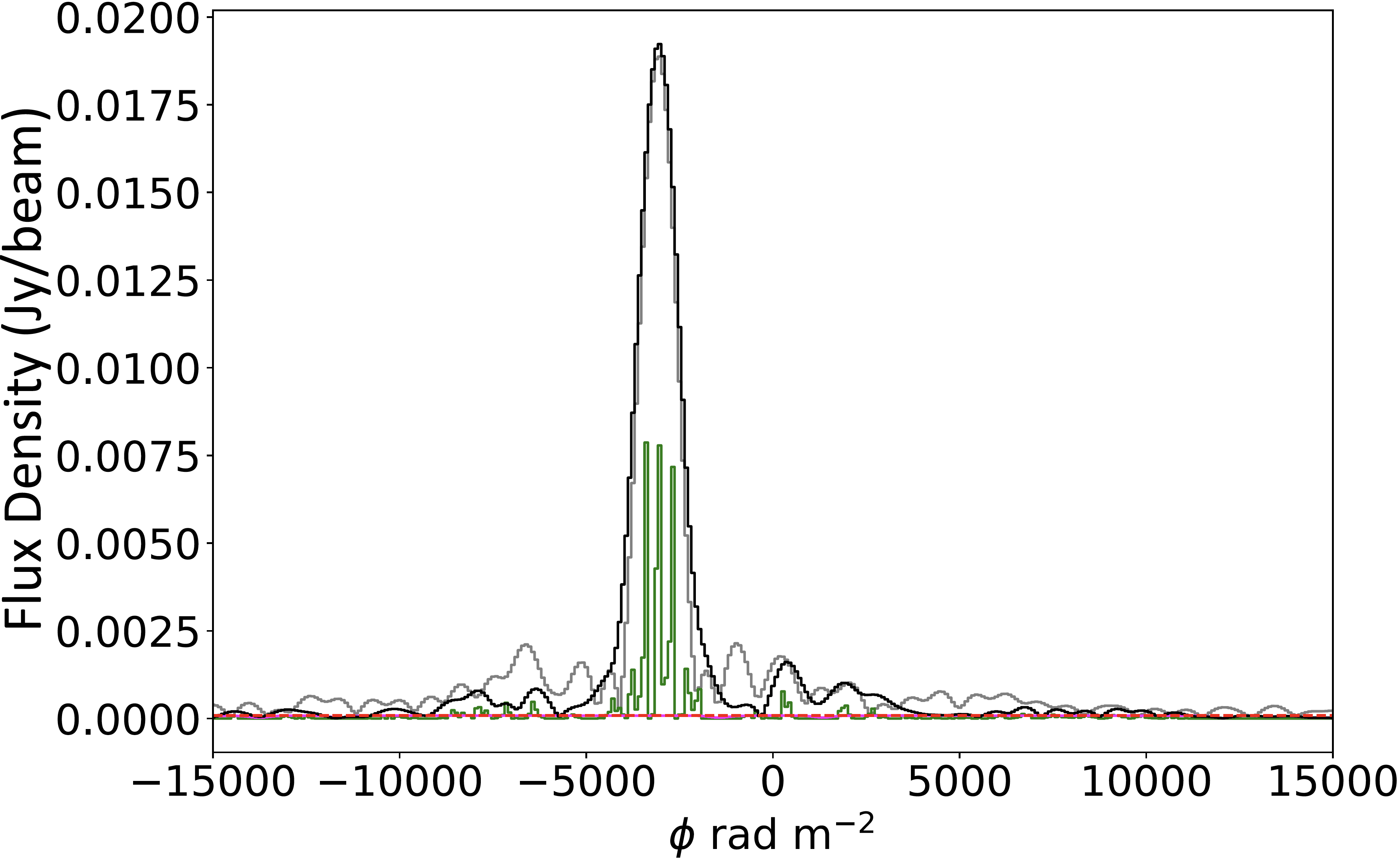}{0.40\textwidth}{5}
    \fig{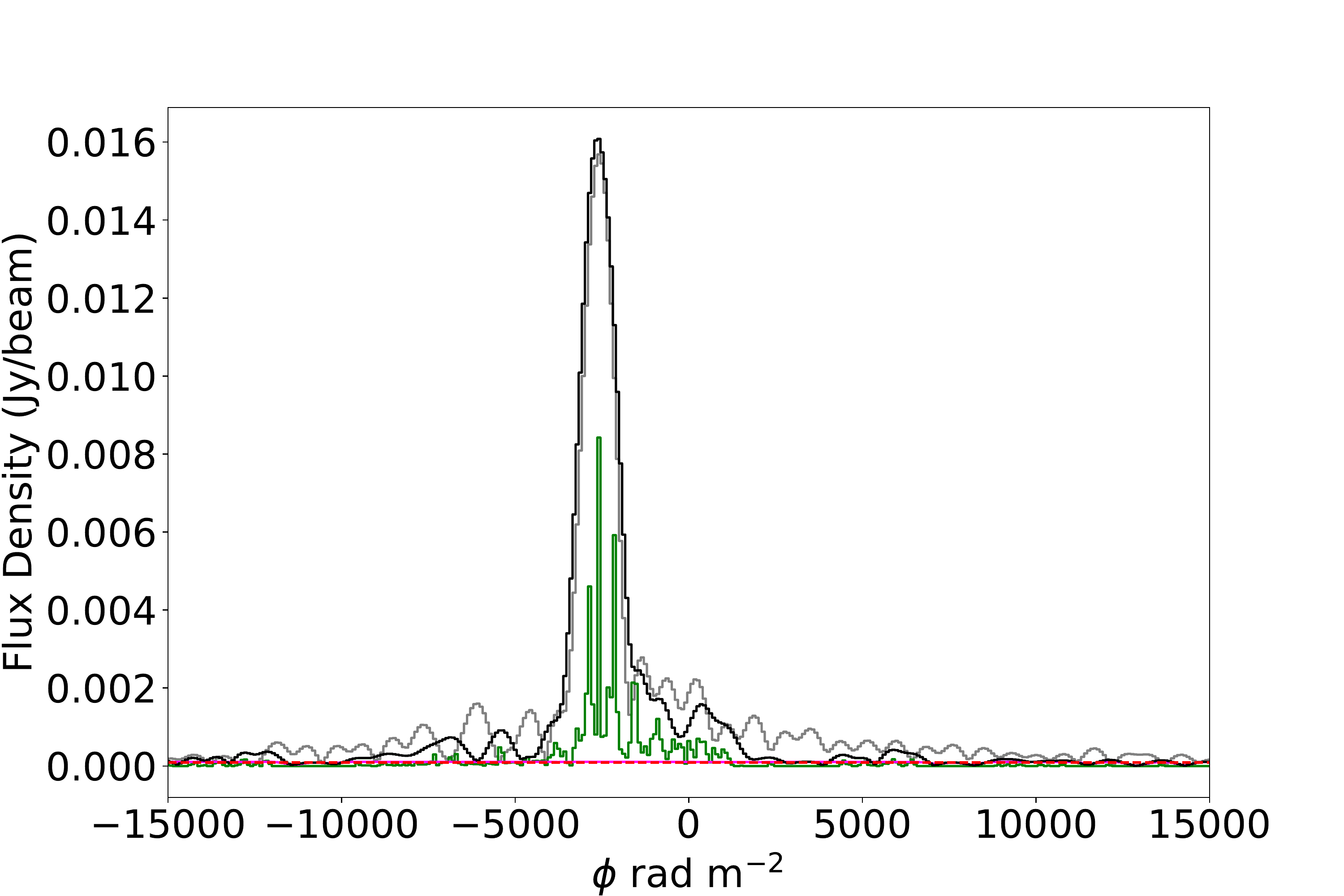}{0.40\textwidth}{6}
    }
    \gridline{\fig{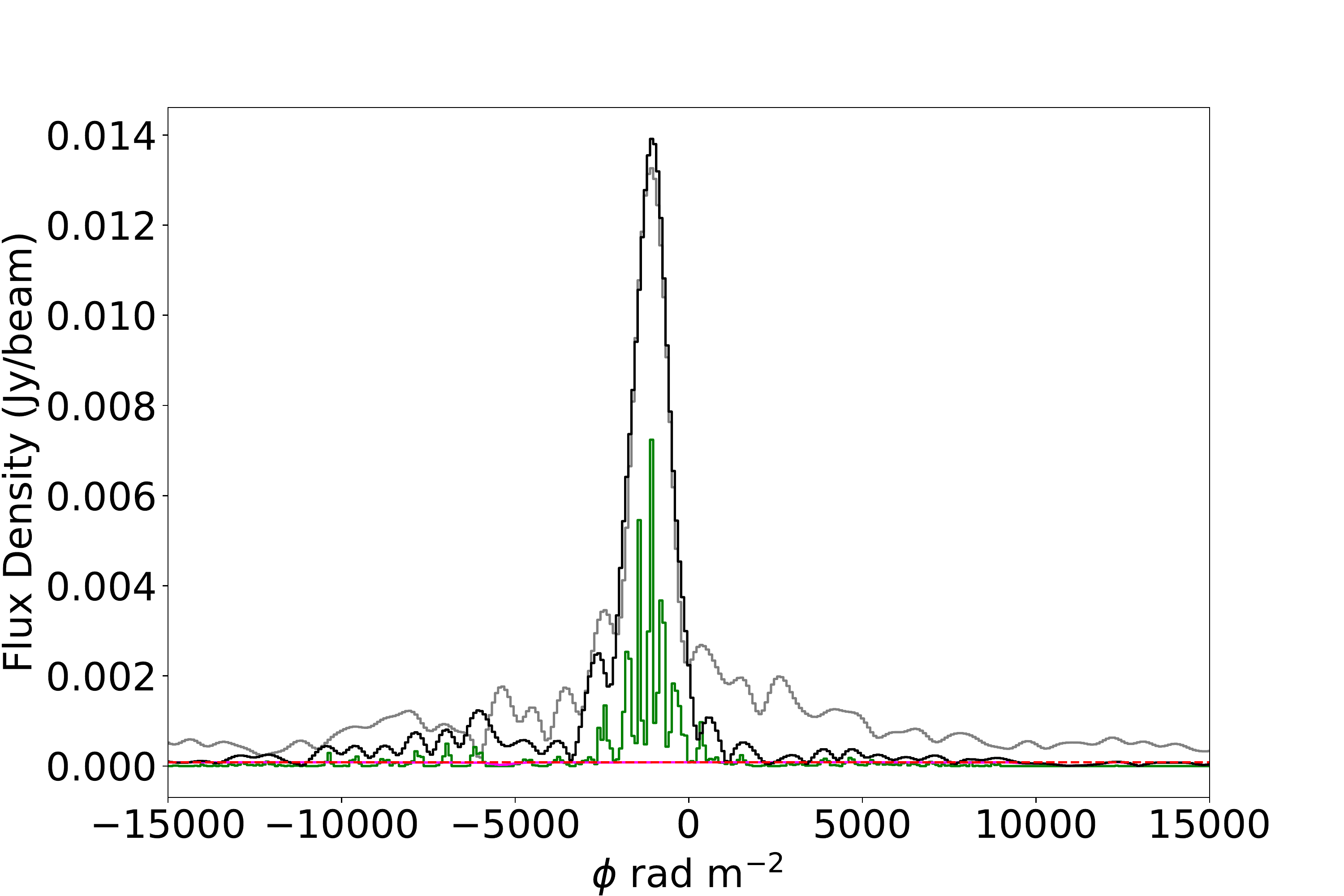}{0.40\textwidth}{7}
    \fig{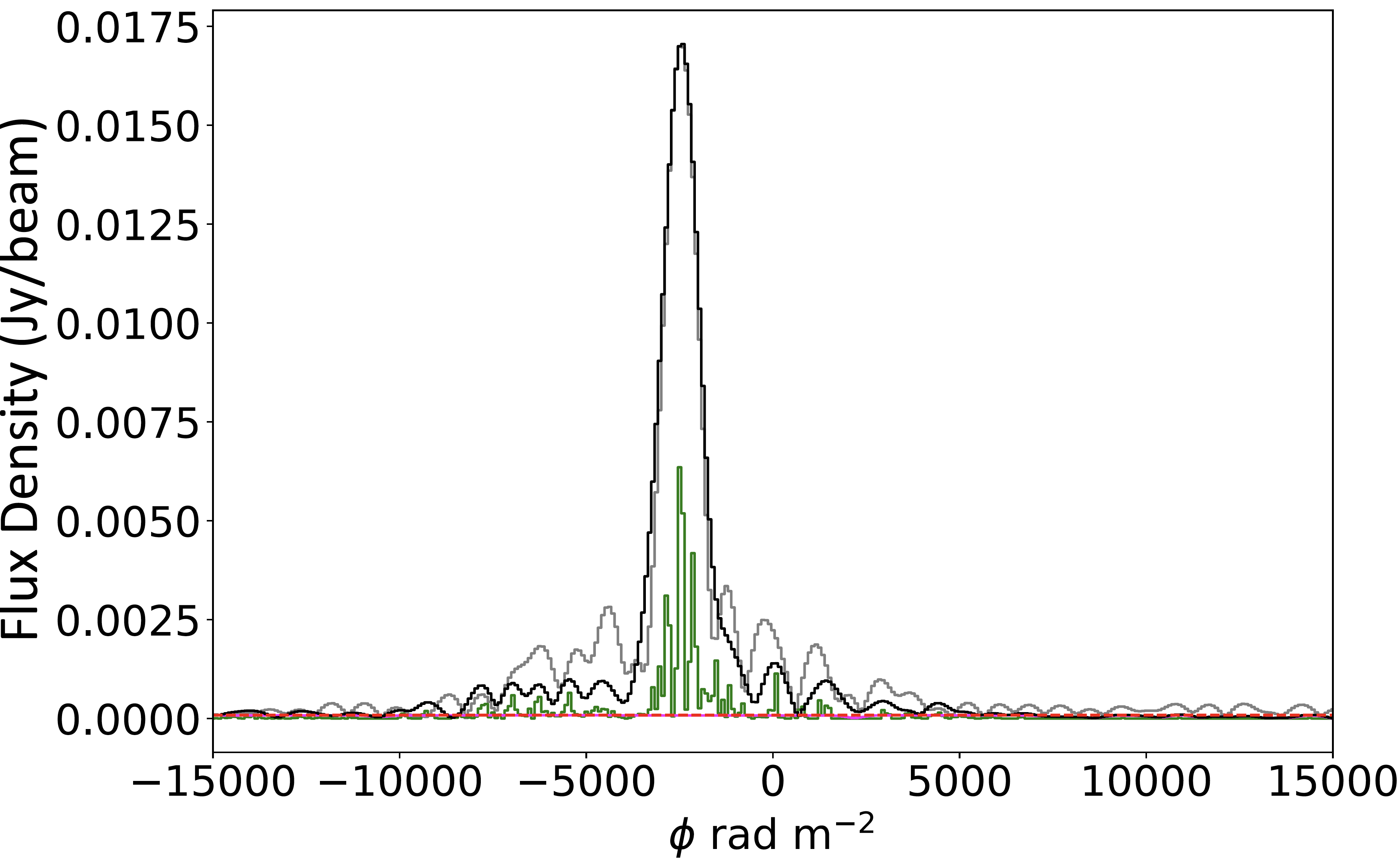}{0.40\textwidth}{8}
    }
    \caption{The FDF spectra obtained for the eight pixels shown in Figure \ref{fig:pix_loc}. For each plot, the label number is the pixel number from Figure \ref{fig:pix_loc} whose FDF is shown. The grey line shows the dirty FDF, the black line shows the cleaned FDF after running RMclean on the spectrum, the green line indicates the locations of clean components, and the red line indicates the cutoff threshold level used for RMclean.}
    \label{fig:FDF}
\end{figure*}
Before starting the Bayesian fitting process, we inspected the cleaned Faraday Dispersion Functions (FDFs) for each of the eight pixels shown in Figure \ref{fig:pix_loc}. These pixels were the focus of the model fitting analysis because they are located in regions of high polarized intensity signal-to-noise ratios. In addition, half of the pixels are in regions of parallel magnetic field and half are in regions of rotated magnetic field as shown in Figure \ref{fig:pix_loc}. The pixels are also spaced throughout the polarized portion of the NTFs. The FDFs show the Faraday depths encountered along the line-of-sight, with each Faraday depth represented as a peak in the FDF as a function of Faraday depth \citep{Frick2011}. The greater the peak in flux density, the more significant the contribution of the medium to the Faraday rotation along the line-of-sight.

\begin{figure}[hb!]
    \centering
    \includegraphics[width=8.5cm]{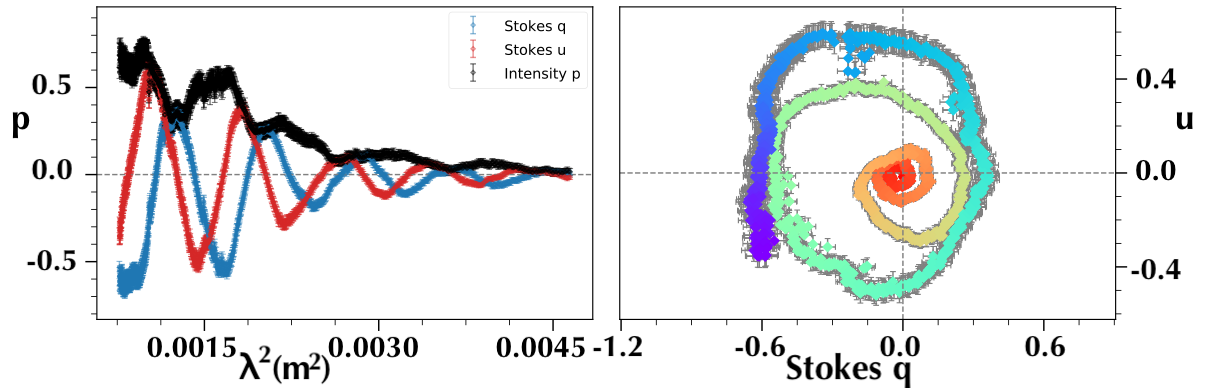}
    \caption{Spectral data for pixel 3 with the fractional q, u, and p data vs wavelength squared shown in the left panel and a plot of q vs u shown in the right panel.}
    \label{fig:case_spec}
\end{figure}

The FDFs are shown in Figure \ref{fig:FDF}, with the x-axis truncated to show only the FDF region with significant flux density peaks. We can see that these pixels have FDFs with one or two significant peaks, indicating that the RM along the lines of sight towards these pixels occurs mostly within one or two media. Smaller flux density peaks possibly correspond to less-significant media components responsible for a fraction of the overall rotation we observe; however, because of their lower flux density these will be less significant components. As a result, we focus on models consisting of only one or two media in this work.

To determine whether models consisting of either one or two media are better fits to the data, we look at an example pixel from the eight pixels shown in Figure \ref{fig:pix_loc}. Figure \ref{fig:case_spec} shows the fractional q, u, and p spectra for pixel 1 in the left panel and the stokes q vs u plot in the right panel. The complexity of the stokes spectra and the stokes q vs u plot indicates that more than one medium component will be needed in our models in order to accurately model the data. We also performed fits using one-medium models consisting of either a single EFD, IFD, or DFR medium and performed Bayesian fitting on all eight pixels using these models. None of these single-medium models was able to fit the data in a convincing manner. We therefore focus on two-media models for this paper.

\subsection{Results of 2-Medium Model Fits}
\begin{deluxetable}{|c|c|c|c|}[ht!]
\tablecaption{Comparison of Reduced $\rm\chi^2$ Values Obtained from Best-Fitting Models}
\tablecolumns{4}
\tablenum{3}
\tablewidth{0pt}
\tablehead{
\colhead{Pix} & \colhead{EFD} & \colhead{DFR} & \colhead{IFD}
}
\startdata
1 & 0.35 & 0.35 & 0.35 \\
2 & 0.49 & 3.09 & 0.49 \\
3 & 0.86 & 3.00 & 2.08 \\
4 & 0.35 & 6.94 & 5.33 \\
5 & 0.30 & 0.30 & 0.30 \\
6 & 10.91 & 0.51 & 0.39 \\
7 & 4.18 & 0.43 & 5.80 \\
8 & 0.45 & 19.91 & 0.41
\enddata
\tablecomments{Pix indicates the pixel number for the model fits as shown in Figure \ref{fig:pix_loc}. EFD, DFR, and IFD indicate the reduced $\rm\chi^2$ values obtained for the best-fit models found using the 2-media component models detailed in the text.}
\label{table:chi_vals}
\end{deluxetable}

\begin{deluxetable}{|c|c|c|c|c|}[ht!]
\tablecaption{Comparison of Evidence Likelihoods Obtained from Best-fitting Models}
\tablecolumns{5}
\tablenum{4}
\tablewidth{0pt}
\tablehead{
\colhead{Pix} & \colhead{B Orientation} & \colhead{EFD} & \colhead{DFR} & \colhead{IFD}
}
\startdata
%1 & \textbf{854.85}\tablenotemark{a} & -746.60 & 195.36 \\
%2 & \textbf{3437.65} & 2133.97 & 2633.34 \\
%3 & \textbf{4579.48} & 2339.43 & 3799.24 \\
%4 & \textbf{-615.69} & -3885.28 & -1099.28 \\
%5 & \textbf{169.48} & -19595.82 & -19257.60 \\
%6 & \textbf{-11797.74} & -20322.31 & -31696.41 \\
%7 & \textbf{1319.95} & -8957.95 & -1410.24 \\
%8 & \textbf{68.99} & -21801.36 & -883.57 
1 & parallel & Max Likely & Unlikely & Likely \\
2 & parallel & Max Likely & Unlikely & Likely \\
3 & parallel & Max Likely & Unlikely & Likely \\
4 & parallel & Max Likely & Unlikely & Likely \\
5 & rotated & Max Likely & Unlikely & Unlikely \\
6 & rotated & Max Likely & Unlikely & Unlikely \\
7 & rotated & Max Likely & Unlikely & Unlikely \\
8 & rotated & Max Likely & Unlikely & Unlikely
\enddata
\tablecomments{Pix indicates the pixel number fit as shown in Figure \ref{fig:pix_loc} and ``B Orientation'' indicates whether the observed magnetic field of the Radio Arc NTFs is parallel or rotated at the pixel location. EFD, DFR, and IFD indicate the likelihood for the different 2-component models discussed in the text.}
\label{table:ev_vals}
\end{deluxetable}

\figsetstart
\figsetnum{10}
\figsettitle{Best-Fitting Corner Plots for All 8 Pixels}
\figsetgrpstart
\figsetgrpnum{10.1}
\figsetgrptitle{Pixel 1}
\figsetplot{Corner_1.png}
\figsetgrpnote{An example corner plot obtained from the best-fitting EFD model for Pixel 1. Each panel displays the final posterior distribution from the Bayesian fitting in greyscale. The parameter values used to generate the best-fitting spectral model are shown in blue.}
\figsetgrpend
\figsetgrpnum{10.2}
\figsetgrptitle{Pixel 2}
\figsetplot{Corner_2.pdf}
\figsetgrpnote{}
\figsetgrpend
\figsetgrpnum{10.3}
\figsetgrptitle{Pixel 3}
\figsetplot{Corner_3.pdf}
\figsetgrpnote{}
\figsetgrpend
\figsetgrpnum{10.4}
\figsetgrptitle{Pixel 4}
\figsetplot{Corner_4.pdf}
\figsetgrpnote{}
\figsetgrpend
\figsetgrpnum{10.5}
\figsetgrptitle{Pixel 5}
\figsetplot{Corner_5.pdf}
\figsetgrpnote{}
\figsetgrpend
\figsetgrpnum{10.6}
\figsetgrptitle{Pixel 6}
\figsetplot{Corner_6.pdf}
\figsetgrpnote{}
\figsetgrpend
\figsetgrpnum{10.7}
\figsetgrptitle{Pixel 7}
\figsetplot{Corner_7.pdf}
\figsetgrpnote{}
\figsetgrpend
\figsetgrpnum{10.8}
\figsetgrptitle{Pixel 8}
\figsetplot{Corner_8.pdf}
\figsetgrpnote{}
\figsetgrpend
\figsetend

\begin{figure*}
    \includegraphics[width=18cm]{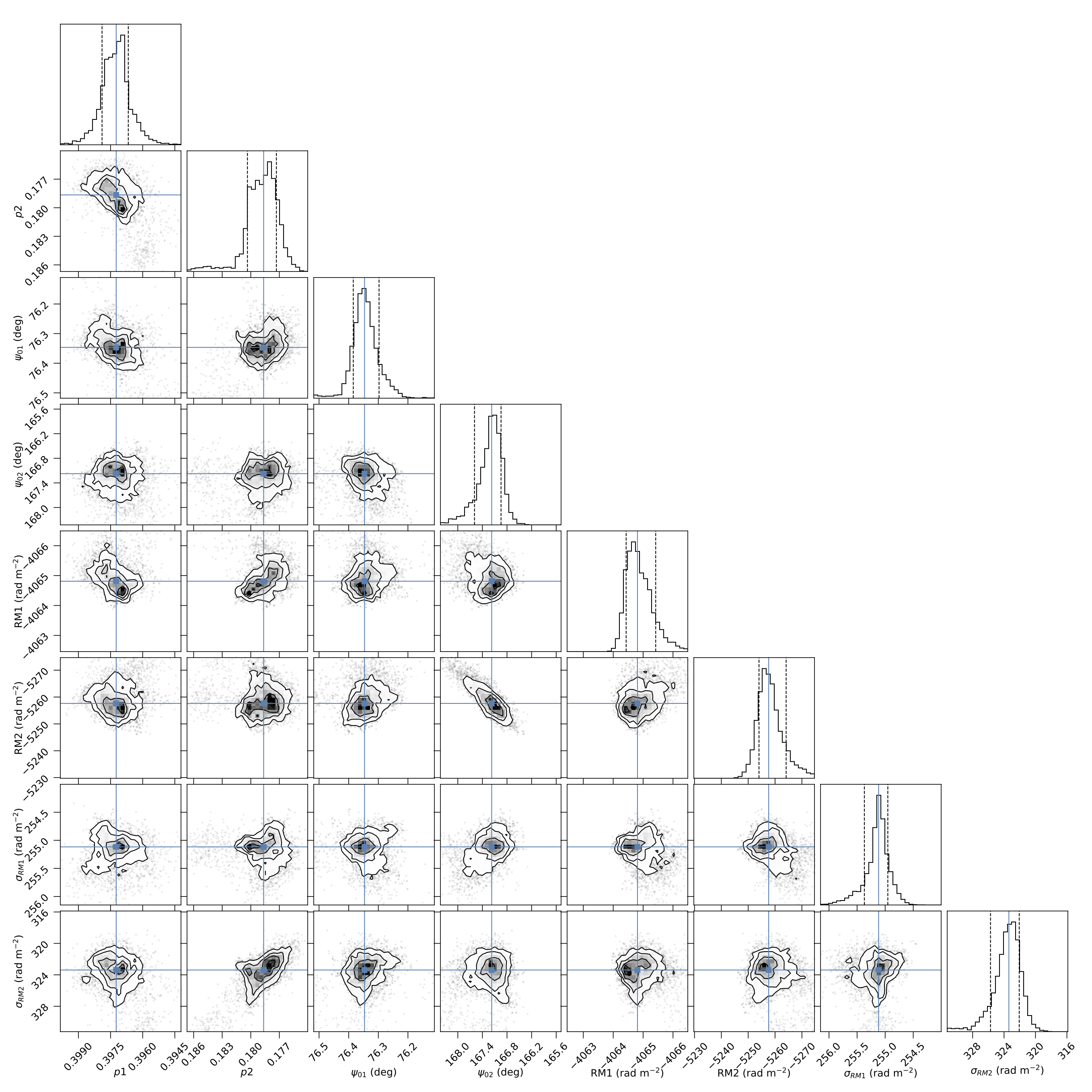}
    \caption{An example corner plot obtained from the best-fitting EFD model for Pixel 1. Each panel displays the final posterior distribution from the Bayesian fitting in greyscale. The parameter values used to generate the best-fitting spectral model are shown in blue. The complete figure set (8 images) is available in the online journal.}
    \label{fig:corner}
\end{figure*}

\figsetstart
\figsetnum{11}
\figsettitle{Best-Fitting Spectral Plots for All 8 Pixels}
\figsetgrpstart
\figsetgrpnum{11.1}
\figsetgrptitle{Pixel 1}
\figsetplot{Spec_1.pdf}
\figsetgrpnote{The spectral plots obtained for the best-fitting EFD model with the highest evidence value for pixel 1. The upper left panel shows the total intensity spectrum for the pixel with the polynomial fit to the data shown as a red line. The upper right panel shows the polarization angle vs wavelength squared with the best fit to the data shown as a red line. The lower two panels show the same data as shown in Figure \ref{fig:case_spec}, with the addition that the best fitting model solutions to the data are shown as solid lines in these panels.}
\figsetgrpend
\figsetgrpnum{11.2}
\figsetgrptitle{Pixel 2}
\figsetplot{Spec_2.pdf}
\figsetgrpnote{}
\figsetgrpend
\figsetgrpnum{11.3}
\figsetgrptitle{Pixel 3}
\figsetplot{Spec_3.pdf}
\figsetgrpnote{}
\figsetgrpend
\figsetgrpnum{11.4}
\figsetgrptitle{Pixel 4}
\figsetplot{Spec_4.pdf}
\figsetgrpnote{}
\figsetgrpend
\figsetgrpnum{11.5}
\figsetgrptitle{Pixel 5}
\figsetplot{Spec_5.pdf}
\figsetgrpnote{}
\figsetgrpend
\figsetgrpnum{11.6}
\figsetgrptitle{Pixel 6}
\figsetplot{Spec_6.pdf}
\figsetgrpnote{}
\figsetgrpend
\figsetgrpnum{11.7}
\figsetgrptitle{Pixel 7}
\figsetplot{Spec_7.pdf}
\figsetgrpnote{}
\figsetgrpend
\figsetgrpnum{11.8}
\figsetgrptitle{Pixel 8}
\figsetplot{Spec_8.pdf}
\figsetgrpnote{}
\figsetgrpend
\figsetend

\begin{figure*}[hb!]
    \includegraphics[width=18cm]{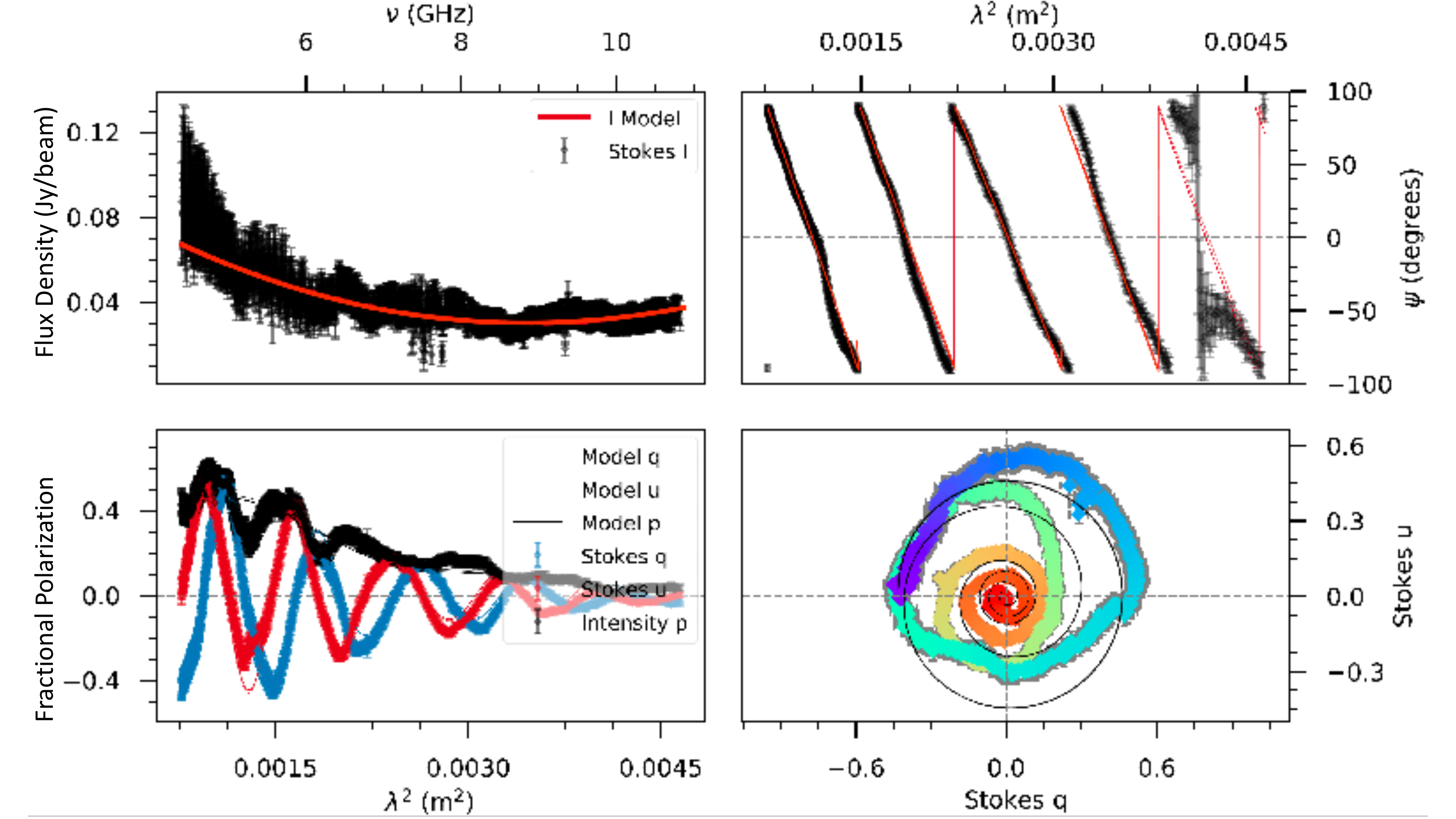}
    \caption{The spectral plots obtained for the best-fitting EFD model with the highest evidence value for pixel 1. The upper left panel shows the total intensity spectrum for the pixel with the polynomial fit to the data shown as a red line. The upper right panel shows the polarization angle vs wavelength squared with the best fit to the data shown as a red line. The lower two panels show the same data as shown in Figure \ref{fig:case_spec}, with the addition that the best fitting model solutions to the data are shown as solid lines in these panels. The complete figure set (8 images) is available in the online journal.}
    \label{fig:spec}
\end{figure*}
\begin{deluxetable*}{|c|c|c|c|c|c|c|c|c|c|}[ht!]
\tablecaption{Parameter Values for the Models with Highest Evidence Values}
\tablecolumns{10}
\tablenum{5}
\tablewidth{0pt}
\tablehead{
\colhead{Pix} & \colhead{B Field} & \colhead{p1} & \colhead{p2} & \colhead{$\rm\chi_1$} & \colhead{$\rm\chi_2$} & \colhead{RM1} & \colhead{RM2} & \colhead{$\rm\sigma_{RM1}$} & \colhead{$\rm\sigma_{RM2}$} \\
\colhead{} & \colhead{Orientation} & \colhead{} & \colhead{} & \colhead{(rad)} & \colhead{(rad)} & \colhead{(rad m$\rm^{-2}$)} & \colhead{(rad m$\rm^{-2}$)} & \colhead{(rad m$\rm^{-2}$)} & \colhead{(rad m$\rm^{-2}$)}
}
\startdata
1 & parallel & 0.23 & 0.41 & 0.03 & 69.9 & -5356 & -4019 & 377 & 260 \\
2 & parallel & 0.70 & 0.53 & 58.4 & 60.1 & -3900 & -3191 & 307 & 438 \\
3* & parallel & 0.17 & 0.66 & 59.2 & 103.2 & 409 & -3936 & 333 & 318 \\
4 & parallel & 0.70 & 0.28 & 89.4 & 156.0 & -3088 & -1726 & 365 & 400 \\
5 & rotated & 0.70 & 0.70 & 28.0 & 101.3 & -2778 & -3316 & 250 & 234 \\
6 & rotated & 0.70 & 0.70 & 91.2 & 38.2 & -2710 & -2160 & 169 & 235 \\
7 & rotated & 0.37 & 0.70 & 143.4 & 68.1 & -1695 & -964 & 411 & 299 \\
8* & rotated & 0.70 & 0.68 & 69.3 & 8.5 & -2434 & -6977 & 219 & 814
\enddata
\tablecomments{Pix indicates the pixel fit as shown in Figure \ref{fig:pix_loc}, B field Orientation indicates the orientation of the observed magnetic field of the Radio Arc NTFs at the location of the pixel studied. p1 and p2 display the fractional polarizations for the 2 external media identified, $\rm\chi_1$ and $\rm\chi_2$ indicate the polarization angles, RM1 and RM2 list the RM values, and $\rm\sigma_{RM1}$ and $\rm\sigma_{RM2}$ indicate the standard deviations of the RM within the media.}
\label{table:param_vals}
\end{deluxetable*}

We tested a suite of 2-media models in order to determine which model best characterizes the behavior observed in the data. The models analyzed were the following: \textbf{1)} a model consisting of 2 EFD media both described by Equation \ref{eq:EFD}, \textbf{2)} a model made of 1 EFD medium and 1 DFR medium constructed using Equation \ref{eq:DFR}, and \textbf{3)} a model using 1 EFD and 1 IFD source quantified by Equation \ref{eq:IFD}. We began by comparing the reduced $\rm\chi^2$ values obtained for each of our 2-component models. These values are shown in Table 3. 

We see in Table 3 that most of our fits have reduced $\rm\chi^2$ values which are either well above or below 1.0. For many of the pixels, the reduced $\rm\chi^2$ values are identical, which makes it challenging to extract any trends in the quality of the model fits. We therefore turn to analyze a parameter known as the evidence to determine which of these models is the most realistic fit to the data. The evidence is a normalization factor that determines the likelihood of the model given the data by using the prior distributions obtained from the Bayesian fitting (Purcell et al. in prep). The evidence statistic serves as a direct way to compare models. A model that describes the data well will have a higher evidence value than a model which describes the data poorly. Therefore, the model which returns the highest evidence value is most representative of the true physical situation producing the rotation of the emission.

Table 4 shows the likelihood of the different models based on the evidence found from RMtools. Columns 3 - 5 indicate the likelihood of the model. The model with the highest evidence was ranked as the most likely model. If a model contained similar evidence to the highest evidence model it was ranked as likely. If a model's evidence was well below the evidence of the most likely model, it was ranked as unlikely. The most likely model for all 8 pixels is the 2 EFD model; however, many pixels could also be described by the 1 IFD and 1 EFD model. However, for all 8 pixels the model consisting of 1 DFR and 1 EFD medium is found to be an unlikely representative of the true physical situation due to its lower evidence.

The parameter values associated with the best-fitting model for each pixel are shown in Table 5. The column information displays, from left to right: the pixel number, the orientation of the magnetic field local to the pixel, the fractional polarization found for the first medium component, the fractional polarization found for the second medium component, the amount of rotation caused by the first medium [deg], the amount of rotation caused by the second medium [deg], the RM value for the first medium [rad m$\rm^{-2}$], the RM value for the second medium [rad m$\rm^{-2}$], the change in RM within the observing beam for the first medium [rad m$\rm^{-2}$], and the change in RM within the observing beam for the second medium [rad m$\rm^{-2}$]. Uncertainties for the parameters were calculated but are not shown in the table because they were found to be at least an order of magnitude of more below the values themselves.

Representative corner and spectral plots obtained for the EFD model yielding the highest evidence value for pixel 1 are shown in Figures \ref{fig:corner} and \ref{fig:spec} respectively. The corner plot shows the posterior distribution after model fitting with each panel showing the likelihood distribution in parameter space. The blue crosshair in each panel indicates the parameter value used for the best-fitting spectrum. Figure \ref{fig:spec} shows the spectrum and best-fit model results for the best-fitting 2 EFD model on the pixel 1 data. It is clear from the spectral fits that even the highest evidence 2-media models do not fully characterize the spectral behavior of the polarization data. Likely, models containing a larger number of media (3 or more media) would be needed to fully characterize these spectra. These additional media would correspond to the lower-flux-density peaks of emission seen in the RMSF spectra shown in Figure \ref{fig:FDF}. 

\section{Discussion} \label{sec:disc}
\subsection{Shell-Like Structures near the Radio Arc NTFs}
We saw in Section \ref{sec:I_res} and Figures \ref{fig:ra_I_1} and \ref{fig:ra_I_2} how the detected helical segments seem to approach a shell of emission encompassing the Radio Arc NTFs. However, it seems that we are only resolving components of this larger shell structure, like the helical segments, without detecting the full shell.

The \citet{Pare2019} VLA observations of the Radio Arc NTFs were also unable to resolve the extended Radio Shell structure. Our ATCA observations, however, have a smaller shortest baseline than the VLA, meaning that we have sensitivity to larger source structures. The shortest VLA baseline of the \citet{Pare2019} observations was $\rm\sim$100 meters, an order of magnitude larger than the shortest baseline in our ATCA observations. While we are not able to fully resolve the Radio Shell in our observations, we are able to observe more of the details of the Radio Shell, as seen in Figures \ref{fig:ra_I_1} and \ref{fig:ra_I_2}.

The shell local to the Radio Arc NTFs was also observed at infrared wavelengths \citep{Egan1998,Levine1999,Price2001,Simpson2007}. \citet{Simpson2007} found the shell to have a diameter of about 10$'$ (23 pc) encompassing the Quintuplet star cluster. \citet{Figer1999} postulated that this shell could be a shock front produced by stellar winds from the Quintuplet star cluster.

In addition, \citet{Butterfield2018} find evidence for an expanding molecular shell in the Radio Arc NTF region (marked with a yellow circle in Figure \ref{fig:ra_I_2}). This shell is similar in morphology to other molecular shells seen in the GC \citep{Tsuboi1997,Oka2001,Tsuboi2009}, and could be produced by out-flowing winds and previous supernova explosions from the Quintuplet cluster. This shell is of smaller scale than the shell seen in our observations or in the \citet{Simpson2007} results, but is embedded within this larger shell. These two shells could therefore have a shared origin -- the out-flowing winds of the Quintuplet star cluster \citep{Butterfield2018}.

Because of our insensitivity to large-scale structures, we cannot verify whether the helical segments are density enhancements of this shell of emission through inspection of the total intensity data. However, through analysis of the depolarization effects experienced along different lines of sight we may be able to determine whether there is a systematic difference between pixels whose emission likely passes through the helical segments compared with those that do not. Such systematic differences may lend support to the theory that these disparate helical segments are in fact part of one larger, cohesive structure. However, this structure may not be a helix. Rather, recent results indicate they could be components of a larger shell-like structure \citep{Pare2019}.

\subsection{Features of Best-Fitting Models}
\subsubsection{Trends in the Model Fits} \label{sec:mod_trends}
To test whether the Radio Shell contributes to the unusual magnetic field pattern for the Radio Arc NTFs seen in \citet{Pare2019}, we can compare the model fits found for pixels 1 - 4 and pixels 5 - 8 to see whether there are any systematic differences between these two sets of pixels. Since pixels 1 - 4 correspond with regions of parallel magnetic field whereas pixels 5 - 8 correspond with regions of rotated magnetic field, as observed by \citet{Pare2019}, any systematic differences in the model fits could indicate that the change in magnetic field orientation is a result of a difference in the Faraday rotation properties along the line-of-sight. To do so, we turn to the trends of the evidence shown in Table 4. These trends reveal that the EFD model is a more likely hypothesis than either the IFD or DFR models. Indeed, the EFD model is the most likely regardless of the orientation of the Radio Arc NTF magnetic field. 

However, the IFD model is also found to be a likely possibility for pixels 1 - 4, whereas this is not the case for pixels 5 - 8. The systematically higher likelihood of the IFD model for pixels 1 - 4 could indicate there are internal Faraday effects along these lines of sight that are either not present or less significant for pixels 5 - 8 as discussed in more detail in Section \ref{sec:shell_like}.

\subsubsection{Nature of Media Identified from Model Fitting}
The universally large RM magnitudes  of $\rm\sim$1000 rad m$\rm^{-2}$ or more (Table 5) indicates that these Faraday media are local to the GC rather than within intervening spiral arms of the Galactic Disk. Such high RM magnitudes are highly unusual outside of the GC, as seen from surveys of RM values obtained for the Galactic disk (e.g. \citealt{Han2018}). 

There are likely lower RM-magnitude media encountered along the line-of-sight that are located within the spiral arms of the Galactic Disk. Since we are only fitting two-component models to our data, however, we do not resolve these lower RM-magnitude media in our fits. Such media are likely to have RM values of only a few 10 to 100 rad m$\rm^{-2}$, negligibly low compared to the higher-magnitude GC media detected in our fits. Our inability to detect these lower-magnitude media does not reflect on our ability to resolve rotation due to the Radio Shell. Since the Radio Shell is located within the GC, we would expect its RM magnitude to be comparable to previous magnitudes observed for other NTFs and the Radio Arc NTFs \citep{Gray1995,Lang1999a,Lang1999b,Pare2019}.

\subsection{Likelihood of the Radio Shell Causing Alternating Magnetic Field Pattern} \label{sec:shell_like}
The trends in the evidence likelihoods discussed in Section \ref{sec:mod_trends} do not provide clear indication of different Faraday environments between pixels 1 - 4 and pixels 5 - 8. If the Radio Shell were significantly affecting the rotation and leading to the regions of rotated magnetic field observed in \citet{Pare2019}, we would expect to see that the pixels in parallel magnetic field regions would be characterized by a different model for depolarization than the pixels in rotated magnetic field regions. However, we see that the depolarization properties for all 8 pixels are best characterized by the model consisting of 2 EFD media. The fact that this model is most likely for all pixels indicates that if the Radio Shell affects the rotation or depolarization of the Radio Arc NTF emission, it is a comparatively minor effect that is not strongly detected by our fits.

The increased likelihood of the IFD model for pixels 1 - 4 vs pixels 5 - 8 is the only systematic difference observed between the two sets of pixels. We would have expected the opposite trend, however, as an increased likelihood of internal effects in pixels 5 - 8 would have corroborated our model for the Radio Shell being responsible for the rotated magnetic field regions. The internal rotation is unlikely to be caused by the Radio Arc NTFs themselves or we would expect all lines of sight to show evidence of internal rotation. 

The internal rotation detected along these lines-of-sight could be due to the Radio Shell; however, there are other GC structures which could be responsible for this rotation as well. \citet{Heywood2019} revealed the presence of large-scale synchrotron `chimneys' that are possible outflows from Sgr A$\rm^*$. The Radio Arc NTFs are located precisely at the northeastern edge of the chimney as can be seen in \citet{Heywood2019} and \citet{Ponti2021}. The chimney is an edge-brightened synchro1tron source, and so could be the source of the internal rotation detected along these lines-of-sight.

This secondary trend of internal rotation being more likely in parallel magnetic field regions could account for a difference between the line-of-sight depolarization effects experienced between pixels 1 - 4 and 5 - 8. The lines-of-sight toward the Radio Arc NTFs are clearly complex, encountering structures like the chimneys and Radio Shell. Therefore, increasing the number of media components used in our model fits may reveal a more significant difference between the parallel and rotated magnetic field regions. We were unable to perform fits using a larger number of media, however, due to computational, statistical, and time limitations.

\subsection{Possible Alternative Mechanisms Producing Alternating Magnetic Field Pattern} \label{sec:alt_mech}
It is unclear from the evidence likelihoods in Table 4 whether the Radio Shell is responsible for the regions of rotated magnetic field in the Radio Arc NTFs. Given this result, it is important to consider whether there are other mechanisms which could be producing the alternating magnetic field observed in \citet{Pare2019}. We explore three scenarios:

\textbf{An intrinsic property of the NTFs:} One possibility is that the alternating magnetic field pattern is an intrinsic property of the NTFs themselves. \citet{Yusef-Zadeh1987a} raise the possibility that the NTFs are twisting around each other. This twisting could produce a rotated magnetic field in the regions where the twisting occurs. However, we see no evidence of twisting in our total intensity data (Figures \ref{fig:ra_I_1} and \ref{fig:ra_I_2}), so it now seems unlikely this is a significant mechanism explaining the rotated magnetic field regions.

Furthermore, when examining other NTFs it is clear that they possess uniformly parallel intrinsic magnetic field distributions \citep{Gray1995,Lang1999a,Lang1999b}. It is unclear, therefore, why the Radio Arc NTFs would differ from all previously studied NTFs. One aspect to consider in this discussion, however, is that all previously studied NTFs have been isolated NTFs consisting of only one or two filaments. The Radio Arc NTFs, by contrast constitute a system of 10 or more individual filaments. The Radio Arc NTFs are also in a more complex region of the GC than the other NTFs studied to date. It is possible that the close proximity of the individual filaments, coupled with the complexity of the region surrounding the Radio Arc NTFs, has caused the originally parallel magnetic field to become sheared and tangled throughout the extent of the Radio Arc NTFs.

One possible way a tangled or complex magnetic field distribution could be achieved in the Radio Arc NTFs is if current flows along the individual filaments. This would lead to an azimuthal component in the magnetic field around each filament. Interaction between these azimuthal components could lead to complex magnetic field configurations between the filaments. \citet{Benford1988} proposed an electrodynamic model for the Radio Arc NTFs wherein current would flow along the filaments of the system. This idea has been explored in more recent publications and is a viable explanation for the nature of the Radio Arc NTFs \citep{Ferriere2009}. However, there are multiple alternative theories regarding the nature of the Radio Arc NTFs (\citealt{Ferriere2009} and references therein), and so it remains uncertain whether current traces the NTFs.

\textbf{Remaining Faraday effects present along line-of-sight:} Alternatively, it is possible that the rotated magnetic field regions are a result of the method used in \citet{Pare2019} to determine the intrinsic magnetic field orientation. Because the VLA data possessed frequency gaps, \citet{Pare2019} were only able to correct for the RM by using a linear fit of the polarization angle as a function of frequency. However, that method is only able to model a single source of rotation for each line-of-sight.

If, for example, multiple media with large RMs are present along the line-of-sight, the linear fitting method would only accurately correct for the sum of these RM contributions. Any remaining uncorrected rotation along the line-of-sight could result in the determination of erroneous intrinsic magnetic field orientations which retain some Faraday effects. Given the large RM values seen for the media detected as shown in Table 5, it is readily apparent that the linear fitting method employed in \citet{Pare2019} to derive the intrinsic magnetic field could be under-characterizing the rotation mechanisms observed along the line-of-sight. 

\textbf{Complicated structures along the line-of-sight:} As mentioned in Section \ref{sec:shell_like}, the two-medium models used to fit the data may not be adequately able to reveal the effects of the Radio Shell. Performing this model-fitting analysis with models consisting of three or more media may yield greater success. The Radio Shell is possibly the structure producing the extended polarized intensity regions \citep{Pare2019}, meaning the Radio Shell could not only be rotating the polarized emission of the Radio Arc NTFs, but also rotating its own polarized intensity emission.

We expect Faraday rotation to be originating from both the GC and the Galactic Disk. Galactic structure studies analyzing the distribution of HII regions reveal the presence of multiple spiral arms between the Earth and the GC (e.g. \citealt{Hou2009}). Significant Faraday media are more likely to occur within these spiral arms than in the interarm regions because of the higher densities within the arms. The higher densities of the GC make it likely for one or more rotating media being located along the line-of-sight as well.

An example of possible rotating media within the GC would be the chimney structures discussed previously in Section \ref{sec:shell_like} and revealed in \citet{Heywood2019}. Another possible medium would be the Radio Shell, with a third possibility being other density-enhanced structures within the GC. To resolve the complexity of these lines-of-sight and extract the properties of these media individually, it would be necessary to obtain data covering a larger radio frequency range. This larger frequency coverage would allow us to fit more model components to the spectra to disentangle contributions arising from the Radio Shell from the contributions (if any) arising from the chimneys.

Recent high frequency studies of the Radio Arc NTFs were conducted using the Atacama Cosmology Telescope (ACT, \citealt{Guan2021}) at 90, 150, 220 GHz and by the QUIET collaboration at 43 and 95 GHz \citep{Ruud2015}. These observations reveal the Radio Arc NTFs in both total and polarized intensity at higher frequencies than in our study. These studies do not observe an alternating magnetic field pattern for the Radio Arc NTFs, but rather one which is uniformly parallel to the NTFs. However, these observations were made with arcminute-scale resolutions compared to the arcsecond-scale resolutions of \citet{Pare2019}. It is possible that the alternating magnetic field pattern observed for the Radio Arc NTFs in \citet{Pare2019} can only be discerned with a arcsecond-scale spatial resolution, and the alternating pattern is undetectable in \citet{Guan2021} and \citet{Ruud2015} due to the much larger beam sizes of those observations. Furthermore, Faraday rotation is less significant at these higher frequencies, since it scales with $\rm\lambda^2$, and so these observations could be observing the polarization emitted by the NTFs without experiencing significant Faraday effects.

These high frequency studies reveal that observations of the Radio Arc NTFs at arcsecond resolutions at 10s to 100s of GHz could be used to expand on the work presented in this paper. A larger frequency study of the polarized intensity of the Radio Arc NTFs ranging from 1 - 100 GHz would be feasible, and would allow improved sensitivity of depolarizing media to improve on the model fitting results presented in this paper.

\section{Conclusion} \label{sec:conc}
We observed the Radio Arc NTFs using the ATCA telescope over a 7 GHz frequency range from $\sim$4 - 11 GHz. These observations enabled a multi-frequency view of the Radio Arc region in both total and polarized intensity. By fitting depolarization and Faraday rotation mechanisms to our spectro-polarimetric data, we were able to assess which Faraday mechanisms are most likely impacting the emission of the Radio Arc NTFs.
We summarize the key results of this study here:

\begin{enumerate}

    \item Images of the Radio Arc NTFs and the surrounding region reveal the familiar filamentary structure observed in other higher-resolution studies. These observations better image the larger-scale and shell-like structures that surround the Radio Arc NTFs then was possible in previous studies of the region (e.g. \citealt{YM1987,Pare2019}. However, these large-scale features are still not fully represented in our observations because of the missing shortest spacings.
    
    \item The polarized intensity is only observed within a confined region of the Radio Arc NTFs, as seen in Figure \ref{fig:ra_P}. The polarization intensity increases with frequency over our observed frequency range, which corresponds with a decrease in $\rm\lambda^2$. This decrease as a function of $\rm\lambda^2$ is a characteristic marker of depolarization. We observe multiple extended polarized intensity features that extend into regions of low total intensity.

    \item Bayesian model fitting of the polarization spectra reveals that the features observed in the data are best fit by external media encountered along the line-of-sight. These media are likely located within the GC because of their large RM magnitudes, as seen in Table 5. However, the polarization spectra exhibit complex structure on small wavelength scales which are not fully modelled using this method.
    
    \item There is an increase in significance of internal Faraday rotation in the parallel magnetic field regions compared to the rotated magnetic field regions. This systematic difference is likely not an effect caused by the Radio Arc NTFs themselves but rather by some other medium located along the line-of-sight, possibly the Radio Shell.
\end{enumerate}

Follow-up work with observations covering a contiguous frequency range of 10s of GHz would facilitate model fits of additional media components. This improved fitting would make it possible to isolate the effects of structures like the Radio Shell and chimney from the contributions of other media located along the line of sight. Fitting the spectra with more media components in this way would deepen our understanding of which medium is responsible for the rotated magnetic field regions observed in the magnetic field distribution of the Radio Arc NTFs.

\acknowledgements We would like to thank Naomi McClure-Griffiths and John Dickey for their assistance in developing the ATCA proposal used to obtain the data used for this project. This material is based upon work supported by the National Science Foundation under Grant No. AST-1615375. D.P. would like to thank the Macquarie University Astrophysics and Astrophotonics Research Centre and the Commonwealth Scientific and Industrial Research Organization (CSIRO) Space and Astronomy for computational and financial support used for the completion of this research. The Australia Telescope Compact Array is part of the Australia Telescope National Facility which is funded by the Australian Government for operation as a National Facility managed by CSIRO.

\software{
    MIRIAD, \citep{Sault1995},
    Astropy \citep{Greenfield2014},
    CASA \citep{McMullin2007},
    LMFIT \citep{Newville2016},
    Matplotlib \citep{Hunter2007}
    }

\bibliographystyle{aasjournal}
\bibliography{astronomy}

\begin{thebibliography}{}
\expandafter\ifx\csname natexlab\endcsname\relax\def\natexlab#1{#1}\fi
\providecommand{\url}[1]{\href{#1}{#1}}

\bibitem[{{Benford}(1988)}]{Benford1988}
{Benford}, G. 1988, \apj, 333, 735

\bibitem[{{Brentjens} \& {de Bruyn}(2005)}]{Brentjens2005}
{Brentjens}, M.~A., \& {de Bruyn}, A.~G. 2005, \aap, 441, 1217

\bibitem[{{Burn}(1966)}]{Burn1966}
{Burn}, B.~J. 1966, \mnras, 133, 67

\bibitem[{{Butterfield} {et~al.}(2018){Butterfield}, {Lang}, {Morris}, {Mills},
  \& {Ott}}]{Butterfield2018}
{Butterfield}, N., {Lang}, C.~C., {Morris}, M., {Mills}, E.~A.~C., \& {Ott}, J.
  2018, \apj, 852, 11

\bibitem[{{Clark}(1980)}]{Clark1980}
{Clark}, B.~G. 1980, \aap, 89, 377

\bibitem[{{Do} {et~al.}(2019){Do}, {Hees}, {Ghez}, {Martinez}, {Chu}, {Jia},
  {Sakai}, {Lu}, {Gautam}, {O'Neil}, {Becklin}, {Morris}, {Matthews},
  {Nishiyama}, {Campbell}, {Chappell}, {Chen}, {Ciurlo}, {Dehghanfar},
  {Gallego-Cano}, {Kerzendorf}, {Lyke}, {Naoz}, {Saida}, {Sch{\"o}del},
  {Takahashi}, {Takamori}, {Witzel}, \& {Wizinowich}}]{Do2019}
{Do}, T., {Hees}, A., {Ghez}, A., {et~al.} 2019, Science, 365, 664

\bibitem[{{Egan} {et~al.}(1998){Egan}, {Shipman}, {Price}, {Carey}, {Clark}, \&
  {Cohen}}]{Egan1998}
{Egan}, M.~P., {Shipman}, R.~F., {Price}, S.~D., {et~al.} 1998, \apjl, 494,
  L199

\bibitem[{{Ferri{\`e}re}(2009)}]{Ferriere2009}
{Ferri{\`e}re}, K. 2009, \aap, 505, 1183

\bibitem[{{Figer} {et~al.}(1999){Figer}, {Kim}, {Morris}, {Serabyn}, {Rich}, \&
  {McLean}}]{Figer1999}
{Figer}, D.~F., {Kim}, S.~S., {Morris}, M., {et~al.} 1999, \apj, 525, 750

\bibitem[{{Frick} {et~al.}(2011){Frick}, {Sokoloff}, {Stepanov}, \&
  {Beck}}]{Frick2011}
{Frick}, P., {Sokoloff}, D., {Stepanov}, R., \& {Beck}, R. 2011, \mnras, 414,
  2540

\bibitem[{{Gravity Collaboration} {et~al.}(2019){Gravity Collaboration},
  {Abuter}, {Amorim}, {Baub{\"o}ck}, {Berger}, {Bonnet}, {Brandner},
  {Cl{\'e}net}, {Coud{\'e} Du Foresto}, {de Zeeuw}, {Dexter}, {Duvert},
  {Eckart}, {Eisenhauer}, {F{\"o}rster Schreiber}, {Garcia}, {Gao}, {Gendron},
  {Genzel}, {Gerhard}, {Gillessen}, {Habibi}, {Haubois}, {Henning}, {Hippler},
  {Horrobin}, {Jim{\'e}nez-Rosales}, {Jocou}, {Kervella}, {Lacour},
  {Lapeyr{\`e}re}, {Le Bouquin}, {L{\'e}na}, {Ott}, {Paumard}, {Perraut},
  {Perrin}, {Pfuhl}, {Rabien}, {Rodriguez Coira}, {Rousset}, {Scheithauer},
  {Sternberg}, {Straub}, {Straubmeier}, {Sturm}, {Tacconi}, {Vincent}, {von
  Fellenberg}, {Waisberg}, {Widmann}, {Wieprecht}, {Wiezorrek}, {Woillez}, \&
  {Yazici}}]{Abuter2019}
{Gravity Collaboration}, {Abuter}, R., {Amorim}, A., {et~al.} 2019, \aap, 625,
  L10

\bibitem[{{Gray} {et~al.}(1995){Gray}, {Nicholls}, {Ekers}, \&
  {Cram}}]{Gray1995}
{Gray}, A.~D., {Nicholls}, J., {Ekers}, R.~D., \& {Cram}, L.~E. 1995, \apj,
  448, 164

\bibitem[{{Greenfield} {et~al.}(2014){Greenfield}, {Tollerud}, {Robitaille}, \&
  {Developers}}]{Greenfield2014}
{Greenfield}, P., {Tollerud}, E.~J., {Robitaille}, T., \& {Developers}, A.
  2014, in American Astronomical Society Meeting Abstracts, Vol. 223, American
  Astronomical Society Meeting Abstracts \#223, 255.24

\bibitem[{{Guan} {et~al.}(2021){Guan}, {Clark}, {Hensley}, {Gallardo}, {Naess},
  {Duell}, {Aiola}, {Atkins}, {Calabrese}, {Choi}, {Cothard}, {Devlin},
  {Duivenvoorden}, {Dunkley}, {D{\"u}nner}, {Ferraro}, {Hasselfield}, {Hughes},
  {Koopman}, {Kosowsky}, {Madhavacheril}, {McMahon}, {Nati}, {Niemack}, {Page},
  {Salatino}, {Schaan}, {Sehgal}, {Sif{\'o}n}, {Staggs}, {Vavagiakis},
  {Wollack}, \& {Xu}}]{Guan2021}
{Guan}, Y., {Clark}, S.~E., {Hensley}, B.~S., {et~al.} 2021, arXiv e-prints,
  arXiv:2105.05267

\bibitem[{{Han} {et~al.}(2018){Han}, {Manchester}, {van Straten}, \&
  {Demorest}}]{Han2018}
{Han}, J.~L., {Manchester}, R.~N., {van Straten}, W., \& {Demorest}, P. 2018,
  \apjs, 234, 11

\bibitem[{{Heywood} {et~al.}(2019){Heywood}, {Camilo}, {Cotton}, {Yusef-Zadeh},
  {Abbott}, {Adam}, {Aldera}, {Bauermeister}, {Booth}, {Botha}, {Botha},
  {Brederode}, {Brits}, {Buchner}, {Burger}, {Chalmers}, {Cheetham}, {de
  Villiers}, {Dikgale-Mahlakoana}, {du Toit}, {Esterhuyse}, {Fanaroff},
  {Foley}, {Fourie}, {Gamatham}, {Goedhart}, {Gounden}, {Hlakola}, {Hoek},
  {Hokwana}, {Horn}, {Horrell}, {Hugo}, {Isaacson}, {Jonas}, {Jordaan},
  {Joubert}, {J{\'o}zsa}, {Julie}, {Kapp}, {Kenyon}, {Kotz{\'e}}, {Kriel},
  {Kusel}, {Lehmensiek}, {Liebenberg}, {Loots}, {Lord}, {Lunsky}, {Macfarlane},
  {Magnus}, {Magozore}, {Mahgoub}, {Main}, {Malan}, {Malgas}, {Manley},
  {Maree}, {Merry}, {Millenaar}, {Mnyandu}, {Moeng}, {Monama}, {Mphego}, {New},
  {Ngcebetsha}, {Oozeer}, {Otto}, {Passmoor}, {Patel}, {Peens-Hough},
  {Perkins}, {Ratcliffe}, {Renil}, {Rust}, {Salie}, {Schwardt}, {Serylak},
  {Siebrits}, {Sirothia}, {Smirnov}, {Sofeya}, {Swart}, {Tasse}, {Taylor},
  {Theron}, {Thorat}, {Tiplady}, {Tshongweni}, {van Balla}, {van der Byl}, {van
  der Merwe}, {van Dyk}, {Van Rooyen}, {Van Tonder}, {Van Wyk}, {Wallace},
  {Welz}, \& {Williams}}]{Heywood2019}
{Heywood}, I., {Camilo}, F., {Cotton}, W.~D., {et~al.} 2019, \nat, 573, 235

\bibitem[{{Hou} {et~al.}(2009){Hou}, {Han}, \& {Shi}}]{Hou2009}
{Hou}, L.~G., {Han}, J.~L., \& {Shi}, W.~B. 2009, \aap, 499, 473

\bibitem[{{Hunter}(2007)}]{Hunter2007}
{Hunter}, J.~D. 2007, Computing in Science and Engineering, 9, 90

\bibitem[{{Inoue} {et~al.}(1989){Inoue}, {Fomalont}, {Tsuboi}, {Yusef-Zadeh},
  {Morris}, {Tabara}, \& {Kato}}]{Inoue1989}
{Inoue}, M., {Fomalont}, E., {Tsuboi}, M., {et~al.} 1989, in IAU Symposium,
  Vol. 136, The Center of the Galaxy, ed. M.~{Morris}, 269

\bibitem[{{Kaczmarek} {et~al.}(2018){Kaczmarek}, {Purcell}, {Gaensler}, {Sun},
  {O'Sullivan}, \& {McClure-Griffiths}}]{Kaczmarek2018}
{Kaczmarek}, J.~F., {Purcell}, C.~R., {Gaensler}, B.~M., {et~al.} 2018, \mnras,
  476, 1596

\bibitem[{{Lang} {et~al.}(1999{\natexlab{a}}){Lang}, {Anantharamaiah},
  {Kassim}, \& {Lazio}}]{Lang1999a}
{Lang}, C.~C., {Anantharamaiah}, K.~R., {Kassim}, N.~E., \& {Lazio}, T.~J.~W.
  1999{\natexlab{a}}, \apjl, 521, L41

\bibitem[{{Lang} {et~al.}(1997){Lang}, {Goss}, \& {Wood}}]{Lang1997}
{Lang}, C.~C., {Goss}, W.~M., \& {Wood}, O.~S. 1997, \apj, 474, 275

\bibitem[{{Lang} {et~al.}(1999{\natexlab{b}}){Lang}, {Morris}, \&
  {Echevarria}}]{Lang1999b}
{Lang}, C.~C., {Morris}, M., \& {Echevarria}, L. 1999{\natexlab{b}}, \apj, 526,
  727

\bibitem[{{Levine} {et~al.}(1999){Levine}, {Morris}, \& {Figer}}]{Levine1999}
{Levine}, D., {Morris}, M., \& {Figer}, D. 1999, in ESA Special Publication,
  Vol. 427, The Universe as Seen by ISO, ed. P.~{Cox} \& M.~{Kessler}, 699

\bibitem[{{McMullin} {et~al.}(2007){McMullin}, {Waters}, {Schiebel}, {Young},
  \& {Golap}}]{McMullin2007}
{McMullin}, J.~P., {Waters}, B., {Schiebel}, D., {Young}, W., \& {Golap}, K.
  2007, in Astronomical Society of the Pacific Conference Series, Vol. 376,
  Astronomical Data Analysis Software and Systems XVI, ed. R.~A. {Shaw},
  F.~{Hill}, \& D.~J. {Bell}, 127

\bibitem[{{Morris}(2007)}]{Morris2007}
{Morris}, M. 2007, ArXiv Astrophysics e-prints, astro-ph/0701050

\bibitem[{{Morris} \& {Serabyn}(1996)}]{Morris1996}
{Morris}, M., \& {Serabyn}, E. 1996, \araa, 34, 645

\bibitem[{{Newville} {et~al.}(2016){Newville}, {Stensitzki}, {Allen}, {Rawlik},
  {Ingargiola}, \& {Nelson}}]{Newville2016}
{Newville}, M., {Stensitzki}, T., {Allen}, D.~B., {et~al.} 2016, {Lmfit:
  Non-Linear Least-Square Minimization and Curve-Fitting for Python},
  Astrophysics Source Code Library, , , ascl:1606.014

\bibitem[{{Oka} {et~al.}(2001){Oka}, {Hasegawa}, {Sato}, {Tsuboi}, \&
  {Miyazaki}}]{Oka2001}
{Oka}, T., {Hasegawa}, T., {Sato}, F., {Tsuboi}, M., \& {Miyazaki}, A. 2001,
  \pasj, 53, 779

\bibitem[{{O'Sullivan} {et~al.}(2018){O'Sullivan}, {Lenc}, {Anderson},
  {Gaensler}, \& {Murphy}}]{OSullivan2018}
{O'Sullivan}, S.~P., {Lenc}, E., {Anderson}, C.~S., {Gaensler}, B.~M., \&
  {Murphy}, T. 2018, \mnras, 475, 4263

\bibitem[{{O'Sullivan} {et~al.}(2017){O'Sullivan}, {Purcell}, {Anderson},
  {Farnes}, {Sun}, \& {Gaensler}}]{OSullivan2017}
{O'Sullivan}, S.~P., {Purcell}, C.~R., {Anderson}, C.~S., {et~al.} 2017,
  \mnras, 469, 4034

\bibitem[{{O'Sullivan} {et~al.}(2012){O'Sullivan}, {Brown}, {Robishaw},
  {Schnitzeler}, {McClure-Griffiths}, {Feain}, {Taylor}, {Gaensler},
  {Landecker}, {Harvey-Smith}, \& {Carretti}}]{OSullivan2012}
{O'Sullivan}, S.~P., {Brown}, S., {Robishaw}, T., {et~al.} 2012, \mnras, 421,
  3300

\bibitem[{{Par{\'e}} {et~al.}(2019){Par{\'e}}, {Lang}, {Morris}, {Moore}, \&
  {Mao}}]{Pare2019}
{Par{\'e}}, D.~M., {Lang}, C.~C., {Morris}, M.~R., {Moore}, H., \& {Mao}, S.~A.
  2019, \apj, 884, 170

\bibitem[{{Ponti} {et~al.}(2021){Ponti}, {Morris}, {Churazov}, {Heywood}, \&
  {Fender}}]{Ponti2021}
{Ponti}, G., {Morris}, M.~R., {Churazov}, E., {Heywood}, I., \& {Fender}, R.~P.
  2021, \aap, 646, A66

\bibitem[{{Price} {et~al.}(2001){Price}, {Egan}, {Carey}, {Mizuno}, \&
  {Kuchar}}]{Price2001}
{Price}, S.~D., {Egan}, M.~P., {Carey}, S.~J., {Mizuno}, D.~R., \& {Kuchar},
  T.~A. 2001, \aj, 121, 2819

\bibitem[{{Ruud} {et~al.}(2015){Ruud}, {Fuskeland}, {Wehus}, {Vidal}, {Araujo},
  {Bischoff}, {Buder}, {Chinone}, {Cleary}, {Dumoulin}, {Kusaka}, {Monsalve},
  {N{\ae}ss}, {Newburgh}, {Reeves}, {Zwart}, {Bronfman}, {Davies}, {Davis},
  {Dickinson}, {Eriksen}, {Gaier}, {Gundersen}, {Hasegawa}, {Hazumi},
  {Huffenberger}, {Jones}, {Lawrence}, {Leitch}, {Limon}, {Miller}, {Pearson},
  {Piccirillo}, {Radford}, {Readhead}, {Samtleben}, {Seiffert}, {Shepherd},
  {Staggs}, {Tajima}, {Thompson}, \& {QUIET Collaboration}}]{Ruud2015}
{Ruud}, T.~M., {Fuskeland}, U., {Wehus}, I.~K., {et~al.} 2015, \apj, 811, 89

\bibitem[{{Sault} {et~al.}(1995){Sault}, {Teuben}, \& {Wright}}]{Sault1995}
{Sault}, R.~J., {Teuben}, P.~J., \& {Wright}, M.~C.~H. 1995, in Astronomical
  Society of the Pacific Conference Series, Vol.~77, Astronomical Data Analysis
  Software and Systems IV, ed. R.~A. {Shaw}, H.~E. {Payne}, \& J.~J.~E.
  {Hayes}, 433

\bibitem[{{Simpson} {et~al.}(2007){Simpson}, {Colgan}, {Cotera}, {Erickson},
  {Hollenbach}, {Kaufman}, \& {Rubin}}]{Simpson2007}
{Simpson}, J.~P., {Colgan}, S. W.~J., {Cotera}, A.~S., {et~al.} 2007, \apj,
  670, 1115

\bibitem[{{Sokoloff} {et~al.}(1998){Sokoloff}, {Bykov}, {Shukurov},
  {Berkhuijsen}, {Beck}, \& {Poezd}}]{Sokoloff1998}
{Sokoloff}, D.~D., {Bykov}, A.~A., {Shukurov}, A., {et~al.} 1998, \mnras, 299,
  189

\bibitem[{{Tsuboi} {et~al.}(1995){Tsuboi}, {Kawabata}, {Kasuga}, {Handa}, \&
  {Kato}}]{Tsuboi1995}
{Tsuboi}, M., {Kawabata}, T., {Kasuga}, T., {Handa}, T., \& {Kato}, T. 1995,
  \pasj, 47, 829

\bibitem[{{Tsuboi} {et~al.}(2009){Tsuboi}, {Miyazaki}, \&
  {Okumura}}]{Tsuboi2009}
{Tsuboi}, M., {Miyazaki}, A., \& {Okumura}, S.~K. 2009, \pasj, 61, 29

\bibitem[{{Tsuboi} {et~al.}(1997){Tsuboi}, {Ukita}, \& {Handa}}]{Tsuboi1997}
{Tsuboi}, M., {Ukita}, N., \& {Handa}, T. 1997, \apj, 481, 263

\bibitem[{{Yusef-Zadeh}(1989)}]{Yusef-Zadeh1989}
{Yusef-Zadeh}, F. 1989, in IAU Symposium, Vol. 136, The Center of the Galaxy,
  ed. M.~{Morris}, 243

\bibitem[{{Yusef-Zadeh} {et~al.}(2004){Yusef-Zadeh}, {Hewitt}, \&
  {Cotton}}]{Yusef-Zadeh2004}
{Yusef-Zadeh}, F., {Hewitt}, J.~W., \& {Cotton}, W. 2004, \apjs, 155, 421

\bibitem[{{Yusef-Zadeh} \& {Morris}(1987{\natexlab{a}})}]{YM1987}
{Yusef-Zadeh}, F., \& {Morris}, M. 1987{\natexlab{a}}, \apj, 322, 721

\bibitem[{{Yusef-Zadeh} \& {Morris}(1987{\natexlab{b}})}]{Yusef-Zadeh1987}
---. 1987{\natexlab{b}}, \aj, 94, 1178

\bibitem[{{Yusef-Zadeh} \& {Morris}(1987{\natexlab{c}})}]{Yusef-Zadeh1987a}
---. 1987{\natexlab{c}}, \apj, 320, 545

\bibitem[{{Yusef-Zadeh} {et~al.}(1984){Yusef-Zadeh}, {Morris}, \&
  {Chance}}]{YMC1984}
{Yusef-Zadeh}, F., {Morris}, M., \& {Chance}, D. 1984, \nat, 310, 557

\bibitem[{{Yusef-Zadeh} {et~al.}(1986){Yusef-Zadeh}, {Morris}, {Slee}, \&
  {Nelson}}]{Yusef-Zadeh1986a}
{Yusef-Zadeh}, F., {Morris}, M., {Slee}, O.~B., \& {Nelson}, G.~J. 1986, \apj,
  310, 689

\bibitem[{{Yusef-Zadeh} {et~al.}(1997){Yusef-Zadeh}, {Wardle}, \&
  {Parastaran}}]{YWP1997}
{Yusef-Zadeh}, F., {Wardle}, M., \& {Parastaran}, P. 1997, \apjl, 475, L119

\end{thebibliography}

\end{document}